\documentclass[
               DIV=15, font size=10.5pt,
                 onecolumn,
               ]{scrartcl}  

\usepackage{color} 
\usepackage{psfrag} 
\usepackage{caption}
\usepackage{hyperref}
\usepackage{amsmath}
\usepackage{amssymb}
\usepackage{graphics}
\usepackage{mathtools}
 \usepackage{pstool} 
\usepackage{todonotes}
\usepackage{gensymb}
\usepackage{bm}
\usepackage{multirow}
\usepackage{subcaption}

\usepackage{floatrow}

\usepackage{textcomp}

\usepackage[square, numbers,sort&compress]{natbib}
 
\usepackage{pgfplots}
\usepackage{pgfplotstable}
\usepackage{filecontents}

\usepackage{tikz}
\usepackage{tikzscale}

\usetikzlibrary{spy}
\usetikzlibrary{snakes,arrows,shapes}    
\usetikzlibrary{spy}
\usetikzlibrary{backgrounds}  
\usetikzlibrary{decorations}
\usetikzlibrary{positioning}
\usetikzlibrary{patterns}
\usetikzlibrary{calc} 

\usepackage{booktabs}   
\usepackage{makecell} 
\usepackage{colortbl}   

\usepackage{xspace}
\usepackage{color}     
\usepackage{setspace}   
 
\usepackage{soul}
\usepackage{ulem}
\usepackage{currfile}
\usepackage{import}

\usepackage{authoraftertitle} 
\usepackage{authblk} 
\usepackage{cleveref}

\pgfplotsset{compat=1.8}

\newcommand{\findmax}[3]{
    \pgfplotstablesort[sort key={#2},sort cmp={float >}]{\sorted}{#1}%
    \pgfplotstablegetelem{0}{#2}\of{\sorted}%
    \let #3=\pgfplotsretval%
}

\pgfplotscreateplotcyclelist{myCycleListBlackAndWhite}
{%
  {black, mark=o},
  {black, mark=square},
  {black, mark=triangle},
  {black, mark=diamond}, 
  {black, mark=pentagon}, 
  {black, mark=*},
  {black, mark=square*},
  {black, mark=triangle*},
  {black, mark=diamond*}, 
  {black, mark=pentagon*}, 
}

\definecolor{darkgreen}{rgb}{0,0.4,0} 
\definecolor{darkbrown}{rgb}{0.5, 0.396, 0.09}

\definecolor{c1}{rgb}{0.0, 0.4196078431372549, 0.6431372549019608}
\definecolor{c2}{rgb}{1.0, 0.5019607843137255, 0.054901960784313725}
\definecolor{c3}{rgb}{0.6705882352941176, 0.6705882352941176,
0.6705882352941176} \definecolor{c}{rgb}{0.34901960784313724, 0.34901960784313724, 0.34901960784313724}
\definecolor{c4}{rgb}{0.37254901960784315, 0.6196078431372549,
0.8196078431372549} \definecolor{c}{rgb}{0.7843137254901961, 0.3215686274509804, 0.0}
\definecolor{c5}{rgb}{0.5372549019607843, 0.5372549019607843,
0.5372549019607843} \definecolor{c}{rgb}{0.6352941176470588, 0.7843137254901961, 0.9254901960784314}
\definecolor{c6}{rgb}{1.0, 0.7372549019607844, 0.4745098039215686}
\definecolor{c7}{rgb}{0.8117647058823529, 0.8117647058823529,
0.8117647058823529}

\pgfplotscreateplotcyclelist{myCycleListColor}
{%
  {blue, mark=o},
  {red, mark=square},
  {darkbrown, mark=triangle},
  {darkgreen, mark=diamond},
  {black, mark=pentagon},
  {blue, mark=*},
  {red, mark=square*},
  {darkbrown, mark=triangle*},
  {darkgreen, mark=diamond*},
  {black, mark=pentagon*},
}

\pgfplotsset{every axis/.append style= 
              {
                font=\small,
                mark size=2,
                line width = 0.1,
                legend style={font=\small, mark size=3, draw=none, fill=none},
                legend cell align=left,
                cycle list name=myCycleListColor,
              }
            }
            
\pgfplotstableset
{
  every odd row/.style={before row={\rowcolor[gray]{0.9}}},
  every head row/.style=
  {
    before row=
    {%
      \toprule
    },
    after row=\midrule
  },
  every last row/.style=
  {
    after row=\bottomrule
  }
}

\newif\ifdrawboundingbox

\usetikzlibrary{external} 
\tikzset{external/system call={pdflatex \tikzexternalcheckshellescape
-halt-on-error -interaction=batchmode -jobname "\image" "\texsource"}} 

\pgfkeys
{ 
  /pgf/number format/.cd,
  fixed,
  set thousands separator={\,},
  1000 sep in fractionals,
}

\newcolumntype{C}[1]{>{\centering\arraybackslash}m{#1}}
\newcolumntype{R}[1]{>{\raggedright\arraybackslash}m{#1}}
\newcolumntype{L}[1]{>{\raggedleft\arraybackslash}m{#1}}

\newcommand{\delete}[1]{\xspace}

\title{A three-field phase-field model for mixed-mode fracture in rock based on experimental determination of the mode II fracture toughness}

\author[1]{L. Hug\thanks{lisa.hug@tum.de, Corresponding Author}}
\author[2]{M. Potten}
\author[2]{G. Stockinger}
\author[2]{K. Thuro}
\author[1]{S. Kollmannsberger}

 \affil[1]{Chair for Computational Modeling and Simulation,
 Technical University of Munich, Arcisstr. 21, 80333 Munich, Germany}
 \affil[2]{Chair of Engineering Geology, Technical University of Munich, Arcisstr. 21, 80333 Munich, Germany}

\newcommand{\publicationDate}{\today}

\usepackage{fancyhdr}
\date{}
\fancypagestyle{plain}{%
  \fancyhf{}%
  \fancyfoot[L]{
  \vspace{-1.25cm} 
  \footnotesize{Preprint submitted to \textit{Journal}  \hfill
  \publicationDate
  \\
  }} }

\crefname{figure}{Fig.}{Fig.}
\crefname{equation}{Eq.}{Eq.}
\crefname{table}{Tab.}{Tab.}

\drawboundingboxtrue
\drawboundingboxfalse 

\newcommand*{\figref}[2][]{%
	\hyperref[{fig:#2}]{%
		Fig.~\ref*{fig:#2}%
		\ifx\\#1\\%
		\else
		\,#1%
		\fi
	}%
}
\definecolor{changes}{RGB}{0,0,0}
\definecolor{changez}{RGB}{0,0,0}
\definecolor{reviewer}{RGB}{0,0,255}
\definecolor{reviewer2}{rgb}{0,0.6,0}
\definecolor{todos}{RGB}{255,0,0}

\newfloatcommand{capbtabbox}{table}[][\FBwidth]

\begin{document}  

\normalem
\maketitle  
  
\vspace{-1.5cm} 
\hrule 
\section*{Abstract}
In this contribution, a novel framework for simulating mixed-mode failure in rock is presented. Based on a hybrid phase-field model for mixed-mode fracture, separate phase-field variables are introduced for tensile (mode I) and shear (mode II) fracture. The resulting three-field problem features separate length scale parameters for mode I and mode II cracks. In contrast to the  classic two-field mixed-mode approaches it can thus account for different tensile and shear strength of rock. The two phase-field equations are implicitly coupled through the degradation of the material in the elastic equation, and the three fields are solved using a staggered iteration scheme. For its validation, the three-field model is calibrated for two types of rock, Solnhofen Limestone and Pfraundorfer Dolostone. To this end, double-edge notched Brazilian disk (DNBD) tests are performed to determine the mode II fracture toughness. The numerical results demonstrate that the proposed phase-field model is able to reproduce the different crack patterns observed in the DNBD tests. A final example of a uniaxial compression test on a rare drill core demonstrates, that the proposed model is able to capture complex, 3D mixed-mode crack patterns when calibrated with the correct mode~I and mode~II fracture toughness.\\[1em]

 \vspace{.2cm} 
\vspace{0.25cm}\\
\noindent \textit{Keywords:} mode II fracture toughness, mixed-mode failure, brittle fracture,  phase-field modeling, Finite Cell Method 
\vspace{0.35cm}
\hrule 
\vspace{0.15cm}
\captionsetup[figure]{labelfont={bf},name={Fig.},labelsep=colon}
\captionsetup[table]{labelfont={bf},name={Tab.},labelsep=colon}
\tableofcontents
\vspace{0.5cm}
\hrule 

\vspace{0.5cm}
\section{Introduction} \label{sec:intro}

Accurate prediction of fracture in rock and rock-like materials is vital for a number of engineering applications, ranging from building processes to deep geothermal applications. In recent years, numerical methods have vastly complemented geo-mechanical testing and can provide important insights into crack initiation and propagation processes. As an alternative to discrete approaches such as XFEM \cite{moes1999finite}, and cohesive zone models \cite{ortiz1999finite}, the phase-field approach to fracture \cite{bourdin2011time, fran1998revisiting} has gained more and more popularity. Due to its elegant way of representing the crack using a smooth and continuous scalar-variable and its formulation as a minimization problem, the phase-field approach facilitates the solution of complex fracture scenarios. In contrast to the aforementioned discrete approaches, crack propagation follows directly from the solution of a partial differential equation without the need for complex remeshing procedures or ad-hoc criteria for crack initiation. Consequently, a wide range of phase-field approaches have been proposed including models for ductile fracture \cite{ulmer2013phase, ambati2015phase}, heterogenenous \cite{hansendorr2018} or anisotropic material \cite{teichtmeister2017phase} and specific materials such as fiber-reinforced concrete \cite{zhang2019phase, aldakheel2021global} and poro-elastic media \cite{zhou2018phase}.\\
For the simulation of fracture in rock it is important to account for the difference in mode I (tensile) and mode II (shear) fracture resistance, as the fracture toughness for mode II fracture is usually higher. The first and most intuitive phase-field approach to capture the mixed-mode behavior in rock was presented by Zhang \cite{zhang2017modification}. Here, different critical energy release rates for mode I and mode II fracture are introduced and the crack driving force is split into two separate parts which correspond to the different crack modes \cite{zhang2017modification}. Bryant and Sun \cite{bryant2018mixed} propose a modification of the mixed-mode approach with consistent kinematics based on the determination of the local crack propagation direction. The approach by Fan \cite{fan2021quasi} extends the splitting method for masonry-like material \cite{freddi2010regularized} to account for mixed-mode behavior by introducing a split into mode I and mode II components based on the local crack direction similar to \cite{bryant2018mixed}. In contrast to the mixed-mode model proposed in \cite{zhang2017modification}, the latter two methods do not suffer from an overestimation of the driving force under pure mode I loading \cite{fan2021quasi}. However, they require the solution of a maximization problem to determine the local crack driving direction. A further drawback of the above mentioned methods is related to the length scale parameter for the phase-field regularization. As the material strength depends on the choice of the length-scale parameter, Tanné \cite{tanne2018crack} suggests to regard the length-scale as a material property and calibrate it with the material's tensile strength. This, however, can only describe the nucleation of mode I cracks. To overcome this problem, \cite{fei2021double} propose a length insensitive multi-phase-field formulation for the simulation of mixed-mode fracture in quasi-brittle materials. Based on the ideas presented in \cite{bleyer2018}, the approach uses two different phase-fields, one for cohesive tensile fracture and one for frictional shear fracture. In the present contribution, we address the length scale problematic by proposing a three-field phase-field model that uses two different length scales, one for mode I and one for mode II failure. The model can be calibrated using the respective tensile and shear strength of the material. The flexible setting is easy to implement and allows for different splits between mode I and mode II components. That way, the three-field model can be tailored for specific applications and to the available computational resources. \\
Proper calibration of the three-field model for different rocks requires their specific and unique mechanical properties. Decisive parameters for the rock's plastic behavior, its elastic properties and tensile strength, can easily be determined by standardized tests, such as the uniaxial compression test \cite{fairhurst1999DraftIS, mutschler2004} and indirect methods like the Brazilian disc test \cite{andreev1991review, bieniawski1978suggested, tan2015brazilian}. Mixed-mode phase-field models also require the critical energy release rates to properly assess mode I and II fracture initiation and propagation. A variety of different tests have been proposed for the determination of the mode I fracture toughness \cite{wei2017experimental, kuruppu2014isrm} and numerous data has been collected for different types of rock. However, data on the mode II fracture toughness is limited. To obtain the true mode II fracture toughness not only the loading of the test specimen has to be in mode II, but the crack initiation has to be driven by shear. Only a limited number of tests for the determination of the mode II fracture toughness have been proposed, including the punch through shear test \cite{backers2002} and the shear box test \cite{rao2003}. Recently, a Double-edge Notched Brazilian Disk (DNBD) test was suggested by Bahrami \cite{bahrami2020}, which features a simple experimental setup, enables the determination of the true mode II fracture toughness, and readily allows for the observation of fracture patterns using high-speed cameras. In the present contribution, we present a full workflow based on the experimental determination of the mode II fracture toughness using DNBD tests. To the authors' knowledge, this is the first mixed-mode model which is calibrated and successfully applied to reproduce both 2- and 3-dimensional, mixed-mode fracture scenarios.\\
The paper is structured as follows. In Section 2, the three-field model and its discretization with the Finite Cell Method is introduced. The DNBD experiments including the computation of the mode II fracture toughness are presented in Section 3, followed by the numerical results in Section 4. Here, the proposed three-field is validated based on the DNBD tests and a complex application example of a uniaxial compression test  after ISRM SM 1979, with determined uniaxial compressive strength and its stress-strain curve, is presented. 
\\[1em]

\section{A three-field phase-field formulation} \label{sec:phasefield-theory}
In this chapter, the theoretical and numerical background of the phase-field formulation are introduced. The proposed phase-field model is a three-field problem based on the ideas presented in \cite{bleyer2018} and \cite{zhang2017modification} with two separate phase-field variables associated to mode I and mode II fracture, respectively.

\subsection{Mathematical formulation}
\begin{figure}[b]
	\centering
	\includegraphics[width=0.98\textwidth]{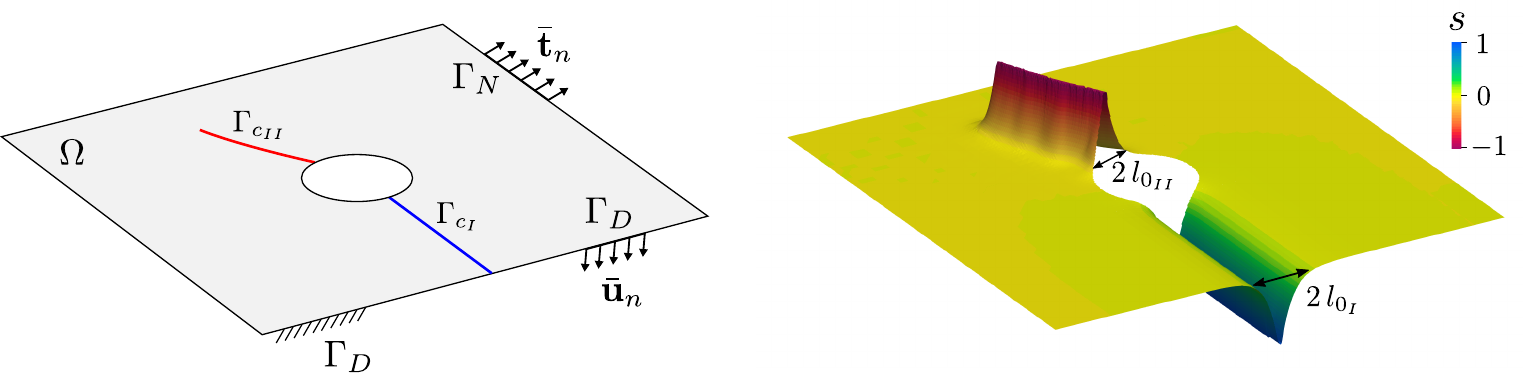}
	\caption{Sharp crack topology (left) and phase-field crack surface (right) adapted from \cite{gerasimov2018penalization}. The discrete cracks $\Gamma_{c_I}$ and $\Gamma_{c_{II}}$ are represented by two distinct phase-field variables $s_I$ and $s_{II}$ ranging from $0$ to $1$. For visualization purposes, these are summarized in a variable $s \in [-1,1]$, which takes a value of $0$ in the undamaged region, $1$ on a tensile crack and $-1$ on a shear crack.  }
	\label{fig:phasefield_setup}
\end{figure}

Let $\Omega \subset \mathcal{R}^d$, $d=2,\,3$ be an open bounded domain which is cut by a set of discrete cracks, as shown in Figure~\ref{fig:phasefield_setup}, left. The set of discrete cracks $\Gamma_{c}$ is split into a set of tensile cracks  $\Gamma_{c_I}$ associated to mode I failure and a set of shear cracks $\Gamma_{c_{II}}$ associated to mode II failure with $\Gamma_{c} = \Gamma_{c_I} \cup\, \Gamma_{c_{II}}$ and $\Gamma_{c_I} \cap\, \Gamma_{c_{II}} = \emptyset$ . The domain boundary $\partial \Omega$ consists of two non-overlapping parts $\Gamma_D$ and $\Gamma_N$ on which Dirichlet and Neumann boundary conditions are prescribed. A point in $\Omega$ is denoted by ${\boldsymbol x}$ and ${\boldsymbol u}({\boldsymbol x}), {\boldsymbol{\varepsilon}}({\boldsymbol x})$ and ${ \boldsymbol{\sigma}}({\boldsymbol x}) \in  \mathcal{R}^d$ are the displacement, strain and stress fields, respectively. An isotropic and linear elastic material with small deformations and quasi-static conditions is assumed. In this case, the strain tensor given as $\boldsymbol{\varepsilon} = \frac{1}{2}\,(\nabla \boldsymbol{u} + \nabla^{\textbf{T}} \boldsymbol{u})$ and the elastic strain density as $\Psi({\boldsymbol{\varepsilon}})=\frac{1}{2}\,\lambda\,\textrm{tr}^2(\boldsymbol{\varepsilon}) + \mu\,\textrm{tr}(\boldsymbol{\varepsilon}^2)$, where $\lambda$ and $\mu$ are the Lam\'e constants.\\

\subsubsection{Background}

The phase-field approach to fracture is based on the variational formulation by Francfort \cite{fran1998revisiting} and its subsequent regularization by Bourdin \cite{bourd2000, bour2008}. Here, the discrete crack is approximated using a scalar variable $s$, the so-called phase-field, which smears the crack over a regularization width $l_0$. The phase-field parameter attains a value of zero on the crack and is one if the material is undamaged. Crack propagation is considered as a minimization problem of the associated functional
\begin{align}
	\label{eq:bourdin_reg_functional}
	E_{l_0}({\textbf u}, s)\,=\, \int_{\Omega}\,g(s)\,\Psi(\boldsymbol{\varepsilon}) \,\textrm{d}{\textbf x}\,+\,\frac{G_{c}}{c_w}\int_{\Omega} \left( \frac{1}{2\,l_0} w(s) \,+\, 2\,l_0 |\nabla s|^2\right) \textrm{d}{\textbf x}\,.
\end{align}
Here, $G_c$ is the critical fracture energy, $g(s)$ is the degradation function which models the loss of stiffness in the material due to damage, $w(s)$ is the energy dissipation function, and $c_w$ a scaling parameter. Formulation (\ref{eq:bourdin_reg_functional}) suffers from interpenetration of crack surfaces and non-physical crack patterns in compression. Thus, commonly an additive split of the elastic strain energy density $\Psi(\boldsymbol \varepsilon) = \Psi^+(\boldsymbol \varepsilon) + \Psi^-(\boldsymbol \varepsilon)$ is used. Based on the spectral split proposed in \cite{miehe2010thermodynamically}, a hybrid formulation was introduced in \cite{ambati2015review}. Here, the split is only accounted for in the phase-field equation, which results in a linear elastic problem. This reduces the computational effort while providing comparable results \cite{ambati2015review}. Following variational theory the Euler-Lagrange equations of the functional \cref{eq:bourdin_reg_functional} can be derived, which yields the following coupled system of equations for the hybrid formulation 
\begin{subequations}
	\label{eq:governeq}
	\begin{align}
	\textrm{div}(\bm{\sigma})\,+\,\rho\,{\bm b}&=0,\qquad  \textrm{where } \bm{\sigma} = g(s)\,\dfrac{\partial \Psi(\boldsymbol  \varepsilon)}{\partial \boldsymbol \varepsilon}\,, \\[0.1cm]
	-4\,l_0^2\,\Delta s\,+\,4\,l_0\,(1\,-\,\eta)\frac{\mathcal{H}}{G_c}\,&=1\,\,.
	\end{align}
\end{subequations}
Here, the AT-2 model with $w(s)=1-s^2$ and $c_w = 1/2$ (\cite{tanne2018crack}) is used. A quadratic degradation function $g(s) = (1-\eta)\,s^2 + \eta$ is chosen, where $\eta$ is a small numerical parameter which prevents full degradation of the material. The coupled system (\ref{eq:governeq}) is subject to the boundary conditions
\begin{align}
\label{eq:bc}
{\textbf u} &= \bar{\textbf{u}}_n\quad &&\textrm{on }\Gamma_{D}, \\
{\boldsymbol{\sigma}}\cdot {\textbf{n}} &= \bar{\textbf{t}}_n &&\textrm{on }\Gamma_N, \\
\nabla d \cdot {\textbf{n}} &= 0 &&\textrm{on }\Gamma_{D} \cup \Gamma_N. 
\end{align}
The history variable $\mathcal{H}$, as introduced by Miehe \cite{miehe2010thermodynamically}, ensures irreversibility of the phase-field and facilitates the use of a staggered solution scheme. It is defined as
\begin{align}
	\mathcal{H}(\textbf{x},t) &\coloneqq \underset{t \in [0,T]}{\textrm{max}}\,\Psi^+(\boldsymbol \varepsilon(\textbf{x},t))\, \nonumber \\
	&=\underset{t \in [0,T]}{\textrm{max}}\, \frac{1}{2}\,\lambda\,\langle\textrm{tr}(\boldsymbol{\varepsilon}(\textbf{x},t))\rangle_{+}^2 + \mu\,\textrm{tr}(\boldsymbol{\varepsilon}_+^2(\textbf{x},t))\,,
\end{align}
where $\langle \,\cdot\, \rangle$ denotes the Macaulay brackets with $\langle\, \cdot\, \rangle_+ = x - |\,x\,|$ and $\boldsymbol{\varepsilon}_+$ is the positive part of the strain tensor resulting from a spectral split. 
In the coupled system (\ref{eq:governeq}), the ratio $\mathcal{H}/G_c$ drives the evolution of the phase-field. Although the critical fracture energies for mode I and mode II fracture can vary considerably, this is not accounted for in the formulation above. To overcome this limitation Zhang \cite{zhang2017modification} proposed a mixed-mode modification of (\ref{eq:governeq}), where two critical fracture energies $G_{c_I}$ and $G_{c_{II}}$ are introduced. The driving force is replaced with a weighted average of mode I and mode II driving forces weighted by their respective critical fracture energies. The phase-field equation (\ref{eq:governeq}a) is modified according to 
\begin{align}
	\label{eq:governeqmixedmode}
	-4\,l_0^2\,\Delta s\,+\,4\,l_0\,(1\,-\,\eta)\left(\frac{\mathcal{H}_{I}}{G_{c_I}}+\frac{\mathcal{H}_{II}}{G_{c_{II}}}\right)\,&=1\,\,,
\end{align}
where the mode I and mode II driving forces $\mathcal{H}_{I}$ and $\mathcal{H}_{II}$ are defined as
\begin{align}
	\label{eq:zhangdrivingforce}
	\mathcal{H}_{I}(\textbf{x},t) &= \underset{t \in [0,T]}{\textrm{max}}\, \frac{1}{2}\,\lambda\,\langle\textrm{tr}(\boldsymbol{\varepsilon}(\textbf{x},t))\rangle_{+}^2\,, \\
	\mathcal{H}_{II}(\textbf{x},t) &= \underset{t \in [0,T]}{\textrm{max}}\, \mu\,\textrm{tr}(\boldsymbol{\varepsilon}_+^2(\textbf{x},t))\,.
\end{align}
The mixed-mode formulation by Zhang \cite{zhang2017modification} uses a single length-scale parameter for both tensile and shear fracture, and thus, is not able to account for different tensile and shear strength of the material. To overcome this limitation we extend the formulation above to a three-field problem, which will be introduced in the next section.

\subsubsection{The three-field phase-field model}
Based on the formulation by Zhang \cite{zhang2017modification}, we propose a three-field phase-field formulation which introduces different scalar variables for tensile and shear failure. The scalar variables  $s_I,\,s_{II} \in [0,1]$ represent mode I and mode II fracture, respectively, with $s_I=0$ on a tensile crack and $s_{II}=0$ on a shear crack as depicted in Figure~\ref{fig:phasefield_setup}. The driving force is split up following \ref{eq:zhangdrivingforce}. Here it is assumed that the tensile phase-field $s_I$ is driven by $\mathcal{H}_{I}$, while the shear field $s_{II}$ is driven by $\mathcal{H}_{II}$. 
Adopting a phase-field evolution according to (\ref{eq:governeq}b) for each of the damage variables the three-field problem is obtained as
\begin{subequations}
	\label{eq:governeqthreefield}
	\begin{align}
		\textrm{div}(\bm{\sigma})\,+\,\rho\,{\bm b}&=0,\qquad  \textrm{where } \bm{\sigma} = g(s_I,\,s_{II}) \mathbb{C}\, {\boldsymbol \varepsilon}\\[0.1cm]
			-4\,l_{0,I}^2\,\Delta s_I\,+\,[4\,l_{0,I}^2\,(1\,-\,\eta)\frac{\mathcal{H}_{I}}{G_{c_I}} + 1]\,s_I\,&=1\,\,,\\
			-4\,l_{0,II}^2\,\Delta s_{II}\,+\,[4\,l_{0,II}^2\,(1\,-\,\eta)\frac{\mathcal{H}_{II}}{G_{c_{II}}} + 1]\,s_{II}\,&=1\,\,.
	\end{align}
\end{subequations}
Here, a degradation function $g(s_I,\,s_{II})$ is defined as
\begin{equation}
g(s_I,\,s_{II}) = (1-\eta)\,(\textrm{min}(s_I, s_{II}))^2 + \eta\,,
\end{equation}
which accounts for the damage of mode I and mode II cracks. Different length scale parameters $l_{0,I}$ and $l_{0,II}$ for shear and tensile fracture, respectively, are introduced. In the case of pure mode I or pure mode II failure, the formulation falls back to the original mixed-mode formulation (\ref{eq:governeqmixedmode}). It should be noted, that the split of the elastic strain energy density in tensile and shear components as proposed by \cite{zhang2017modification} suffers from an overestimation of the force response under pure mode I loading \cite{zhang2017modification}. Alternatively, different approaches based on the split by Amor \cite{amor2009regularized} or directional dependent splits based on local crack coordinates as proposed by Strobl \cite{strobl2015novel} or Steinke \cite{steinke2019phase} can be integrated in the proposed three-field formulation. An overview of existing splitting methods can be found in \cite{fan2021quasi}. In the following, we 

\subsection{Discretization}

The numerical framework is based on the approach presented in \cite{nagaraja2019phase, hug2020}, which combines the phase-field approach with an embedded domain technique, the finite cell method \cite{parvizian2007finite}, and multi-level $hp$-adaptive refinement \cite{zander2015multi}.
\begin{figure}[b]
	\centering
	\includegraphics[width=0.75\textwidth]{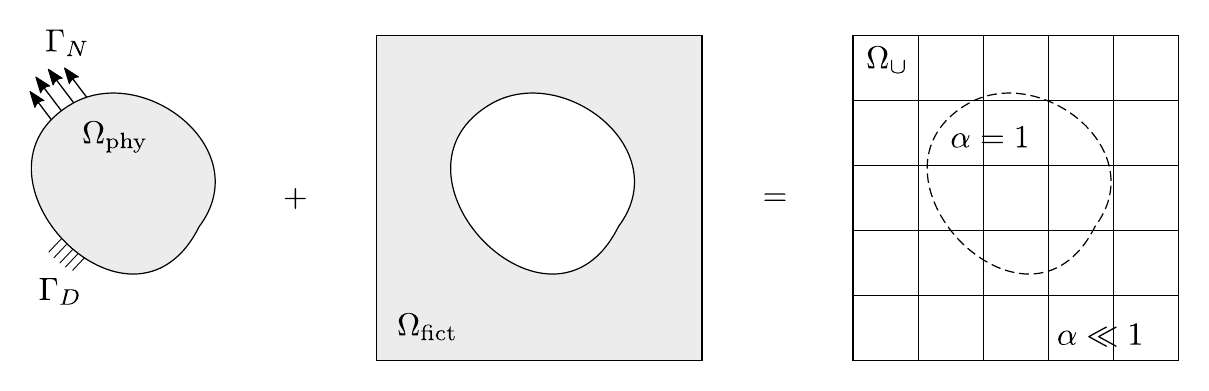}
	\caption{Embedding concept of the Finite Cell Method following \cite{parvizian2007finite}.}
	\label{fig:fcm}
\end{figure}

\subsubsection{The Finite Cell Method}
The finite cell method (FCM) is based on an implicit representation of the geometry. Instead of generating a boundary conforming mesh, the actual geometry is recovered during integration with the help of an indicator function. As depicted in Figure~\ref{fig:fcm}, the physical domain $\Omega_{phy}$ is embedded into a larger domain of simple shape $\Omega_{\cup} = \Omega_{phy} \cup \Omega_{fict}$ which can easily be meshed. To account for the actual geometry, an indicator function $\alpha({\boldsymbol x})$ is defined which takes a value close to zero in the surrounding, so-called fictitious domain and is equal to one in the physical region:
\begin{equation}
\alpha({\boldsymbol x})\,=\,\begin{cases} 1.0\,, \quad \forall {\boldsymbol x} \in \Omega_{phy}, \\ \tilde{\alpha}\,, \quad \forall {\boldsymbol x} \in \textcolor{changes}{\Omega_{fict}}. \end{cases}
\end{equation} 
Here, $\tilde{\alpha}$ is a small numerical parameter greater than but unequal to zero to ensure stability \cite{parvizian2007finite}. The weak form is multiplied by $\alpha({\boldsymbol x})$ eliminating contributions from the fictitious domain. Advanced integration schemes such as quad- and octree-subdivision approaches are needed for a sufficiently accurate integration of the cells cut by the domain boundary \cite{duster2008finite}. For further elaboration on the FCM and its combination with multi-level $hp$-adaptive refinement the reader is referred to \cite{duster2008finite} and \cite{zander2015multi}.

\subsubsection{Weak Form}
Let the trial spaces for the displacement solution $\boldsymbol{\mathcal{S}}_u$, the mode I phase-field solution $\mathcal{S}_{s_{I}}$ and the mode II phase-field solution $\mathcal{S}_{s_{II}}$ be defined as
\begin{align}
\boldsymbol{\mathcal{S}}_u &= \{ \boldsymbol u\, :\, u_i\,\in\,H^1(\Omega),\, u_i|_{\Gamma_{D}} = \bar{u}_i\}\,,\\
\mathcal{S}_{s_{I}} &= \{ s_{I}\, :\, s_{I}\,\in\,H^1(\Omega)\}\,,\\
\mathcal{S}_{s_{II}} &= \{ s_{II}\, :\, s_{II}\,\in\,H^1(\Omega)\}\,,
\end{align}
where $H^1$ refers to the Sobolev space of degree one. Furthermore, let the spaces for the test functions be defined as 
\begin{align}
	\boldsymbol{\mathcal{V}}_u &= \{ \boldsymbol w\, :\, w_i\,\in\,H^1(\Omega),\, w_i|_{\Gamma_{D}} = 0 \}\,,\\
	\mathcal{V}_{s_{I}} &= \{ q_{I}\, :\, q_{I}\,\in\,H^1(\Omega)\}\,,\\
	\mathcal{V}_{s_{II}} &= \{ q_{II}\, :\, q_{II}\,\in\,H^1(\Omega)\}\,.
\end{align}
The weak formulation of the coupled three-field problem for the FCM states:\\ Find $\boldsymbol u \in \boldsymbol{\mathcal{S}}_u $, $s_I \in \mathcal{S}_{s_{I}}$ and $s_{II} \in \mathcal{S}_{s_{II}}$ such that
\begin{subequations}
	\label{eq:weak-galerkin}
	\begin{equation}
	\label{eq:weak-galerkin-el}
	\begin{aligned}
	\left(\boldsymbol{\sigma}, \nabla \boldsymbol{w}\right)_{\Omega_{phy}} \,&+\, (\tilde{\alpha} \,\boldsymbol{\sigma}, \nabla \boldsymbol{w})_{\Omega_{fict}} \,+\, (\beta \, \boldsymbol{u}, \boldsymbol{w})_{\Gamma_{D}}\\[2mm] &= 
	 (\rho\,\boldsymbol{b}, \boldsymbol{w})_{\Omega_{phy}}\, +\, (\boldsymbol{h},\boldsymbol{w})_{\Gamma_{D}}\, +\, (\beta \, \boldsymbol{g}, \boldsymbol{w})_{\Gamma_{D}}\,, \qquad \forall\, \boldsymbol w \in 	\boldsymbol{\mathcal{V}}_u
	\end{aligned}
	\end{equation}
	\begin{equation}
	\label{eq:weak-galerkin-p1}
	\begin{aligned}
	\left(\left[{4\,l_{0,I}}(1-\eta)\frac{\mathcal{H}_{I}}{G_{c_{I}}}+1\right]\,s_{I},q_I\right)_{\Omega_{phy}} &+ \left(\tilde{\alpha} \left[ {4\,l_{0,I}}(1-\eta)\frac{\mathcal{H}_{I}}{G_{c_{I}}}+1\right]\,s_{I},q_{I}\right)_{\Omega_{fict}}\\[2mm]  
	+ \left(4\,l_{0,I}^2\,\nabla s_{I}, \nabla q_{I} \right)_{\Omega_{phy}}  
	&+ \left(\tilde{\alpha}\, 4\,l_{0,I}^2\,\nabla s_{I}, \nabla q_{I} \right)_{\Omega_{fict}} 
	= (1,q_{I})_{\Omega_{phy}}\,, \quad \forall\, q_I \in \mathcal{V}_{s_{I}}
	\end{aligned}
	\end{equation}
	\begin{equation}
	\label{eq:weak-galerkin-p2}
	\begin{aligned}
	\left(\left[ {4\,l_{0,II}}(1-\eta)\frac{\mathcal{H}_{II}}{G_{c_{II}}}+1\right]\,s_{II},q_{II}\right)_{\Omega_{phy}} &+ \left(\tilde{\alpha} \left[ {4\,l_{0,II}}(1-\eta)\frac{\mathcal{H}_{II}}{G_{c_{II}}}+1\right]\,s_{II},q_{II}\right)_{\Omega_{fict}}\\[2mm]    
	+ \left(4\,l_{0,II}^2\,\nabla s_{II}, \nabla q_{II} \right)_{\Omega_{phy}}
	&+ \left(\tilde{\alpha}\, 4\,l_{0,II}^2\,\nabla s_{II}, \nabla q_{II} \right)_{\Omega_{fict}} 
	= (1,q_{II})_{\Omega_{phy}}\,, \quad \forall\, q_{II} \in \mathcal{V}_{s_{II}}
	\end{aligned}
	\end{equation}
\end{subequations}
In (\ref{eq:weak-galerkin-el}), the penalty method is used to apply Dirichlet boundary conditions for the elastic problem in a weak sense, where $\beta$ is the penalty parameter.

\subsubsection{Solution of the Coupled Three-Field Problem}
The system (\ref{eq:weak-galerkin}) is discretized in a finite element setting using integrated Legendre polynomials as basis functions for
the finite test and trial spaces as explained in \cite{nagaraja2019phase}. In each displacement step of the quasi-static simulation, a staggered solution scheme is used to solve the discretized equations. First Eq. (\ref{eq:weak-galerkin-p1}) is solved for the tensile field, then Eq. (\ref{eq:weak-galerkin-p2}) is solved for the shear field and finally Eq. (\ref{eq:weak-galerkin-el}) is solved for the displacements $\boldsymbol u$, which are used to update the history variables $\mathcal{H}_{I}$ and $\mathcal{H}_{II}$. In the next staggered step, the phase-field equations are solved using the updated history variables and the staggered iterative scheme continues until convergence. As a stopping criterion for the staggered procedure, the residua of the three solution fields are compared against a certain threshold. The iterations are terminated after staggered step $i$ if
\begin{equation}
s_\textrm{tol} < \varepsilon\,, \quad \textrm{ where } s_\textrm{tol, i} = \textrm{max}(\mathcal{R}_{\boldsymbol u}, \mathcal{R}_{s_I}, \mathcal{R}_{s_{II}}  ).
\end{equation}
In contrast to the classic two-field mixed-mode approaches and similar to Fei \cite{fei2021double}, the history variables $\mathcal{H}_{I}$ and $\mathcal{H}_{II}$ are updated depending on the dominating crack mode. If $\mathcal{H}_{I} \geq \mathcal{H}_{II}$ or $s_{I}({\boldsymbol x}) < 0.5 $ we assume that a mode I is present and only perform an update $\mathcal{H}_{I}$. Accordingly, if $\mathcal{H}_{I} < \mathcal{H}_{II}$ or $s_{II}({\boldsymbol x}) < 0.5 $, we assume that a mode II crack is present and update $\mathcal{H}_{II}$.

\section{Determination of the mode II fracture toughness} \label{sec:experiments}
In this section, the mode II fracture toughness is determined for two types of rock, namely Solnhofen Limestone (SPK) and Pfraundorfer Dolostone (PFD). The expected crack path is shown in \mbox{Figure~\ref{fig:setup_bahrami}b)}, and consists of a straight shear crack connecting the two external notches. The two investigated rocks are analogue geothermal carbonate reservoir rocks. The Solnhofer limestone is very homgenous and very fine grained ($0.1$ - $0.055$ mm) and the Pfraundorfer Dolostone consists of $99\%$ dolomite with small vugs \cite{thuro2019}. The experimental setup is based on the double-edge notched Brazilian disk (DNBD) tests presented in \cite{bahrami2020}. In contrast to conventional mode II tests, the DNBD test features not only shear-based crack tip loading, but also the material failure is shear-induced. 

\subsection{Double-edge Notched Brazilian Disk (DNBD) tests}

\begin{figure}[t!]
	\centering
	\includegraphics[width=0.85\textwidth]{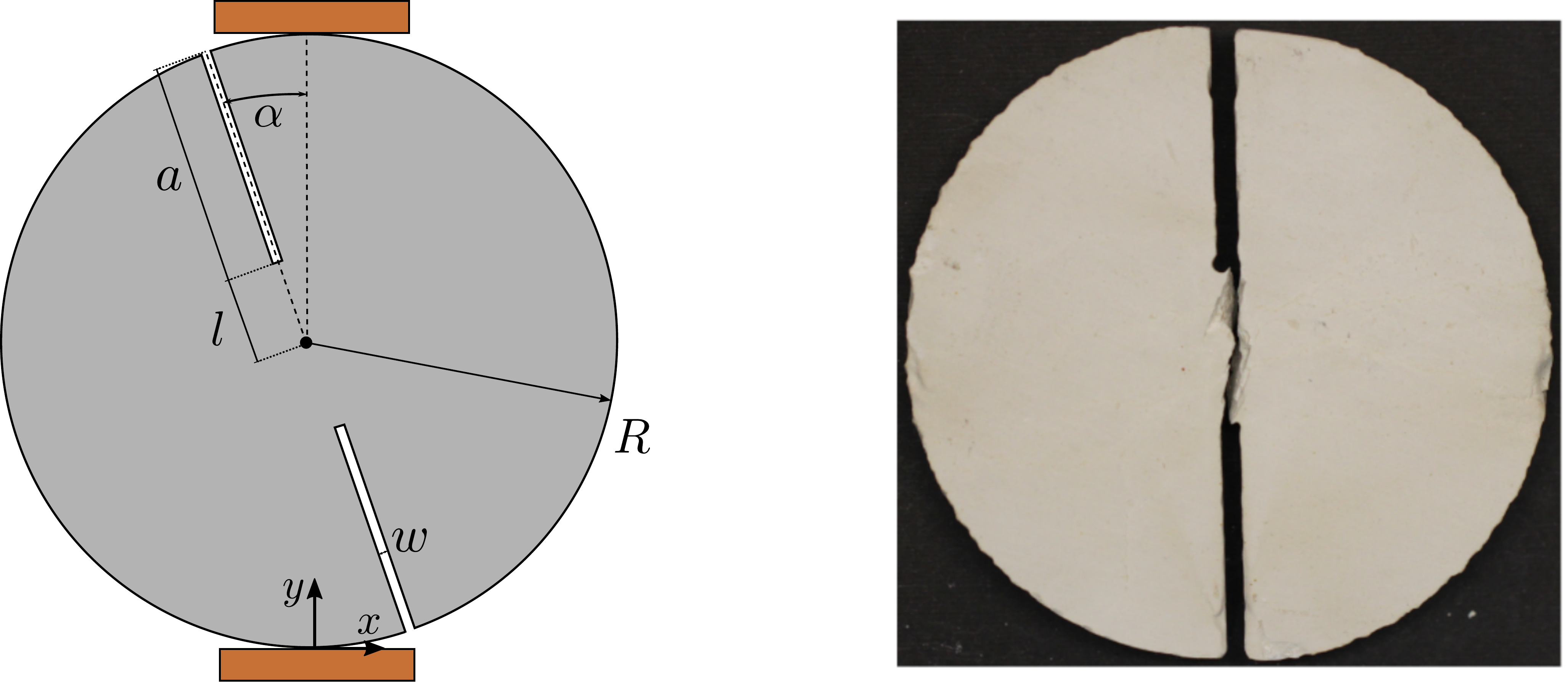}
	\caption{Schematic setup for determination of the mode II fracture thoughness (left), and expected crack pattern (right) following \cite{bahrami2020} . }
	\label{fig:setup_bahrami}
\end{figure}

The experimental setup is schematically depicted in Figure~\ref{fig:setup_bahrami} a). A Brazilian disk specimen containing two external notches of length $a$ and thickness $w$ is diametrically compressed at an angle $\alpha$. Wooden plates are attached to the top and bottom of the specimen to induce the load. The use of flexible materials allows for an even distribution of the load and prevents local concentration of the stress, which can lead to fracturing at the loading points. At the same time it introduces an additional non-linear displacement. Increasing the load angle $\alpha$, the mode I intensity factor $K_I$ increases, which results in smaller shear stresses and consequently a higher failure load. Bahrami~\cite{bahrami2020} suggest to use an angle $\alpha$ in the range of $10\degree$ to $20\degree$ to prevent failure at the loading points, which can occur for very small as well as too large loading angles. Moreover, they state that the size of the ligament $l$ which defines the distance between the two notches (cf. Figure~\ref{fig:setup_bahrami}) should be chosen depending on the disk radius such that $0.2<l/R<0.35$.

\subsection{Experimental results}

Brazilian Disk samples of SPK and PFD with a radius of $47$ mm and a thickness of $20$ mm were prepared by cutting two external notches of length $37$ mm. This results in a ligament of $20$ mm and a ligament to radius ratio of $l/R = 0.21$, which lies within the suggested optimal range \cite{bahrami2020}. Two different geometries were tested, one with notches of width $2.2$ mm which were inserted by hand using a saw and a second geometry with notches of width $1.1$ mm cut with a water jet cutter. While the water jet cutter resulted in perfectly aligned notches, the samples prepared by hand showed slight deviations in the geometry including misaligned notches and different notch length. For all experiments, a loading angle of $\alpha = 15 \degree$ is chosen. The specimen are loaded until fracture with displacement controlled steps of $0.5$ mm/min. The fracture process is recorded using a Photron Mini UX100 high-speed 10000 fps camera with a resolution of $1280 \times 480$ pixels. A spray pattern is applied to the specimen to perform a digital image correlation (DIC) analysis using the open-source 2D-DIC software Ncorr \cite{blaber2021}.

\subsubsection{Calculation of fracture toughness}

\begin{table}[t]
	\RawFloats
	
	\parbox[t][][t]{.5\linewidth}{
		\centering
		\setlength{\tabcolsep}{7pt}
		\renewcommand{\arraystretch}{1.2}
		\begin{tabular}[t]{c|c|c|c}
			$a/R $ & $\alpha $ & $K_I^{\ast}$ & $K_{II}^{\ast}$ \\ \hline
			$0.75$ & $ 15 \degree$&$	-0.94 $ & $-2.38$ \\
			$0.80$ & $ 15 \degree$&$	-0.88 $ & $-2.52$ \\
			$0.85$ & $ 15 \degree$&$	-0.88 $ & $-2.78$ \\
		\end{tabular} 
		\caption{Normalized crack tip parameters for the relevant geometries extracted from  \cite{bahrami2020}.}
		\label{tab:GcEvaluation}
	}
	\hfill
	\parbox[t][][t]{.45\linewidth}{
		\centering
		\setlength{\tabcolsep}{7pt}
		\renewcommand{\arraystretch}{1.2}
		\begin{tabular}[t]{l|c|c|c|c}
			&$K_{I} $ & $K_{II} $ & $F_\textrm{max}$&$	K_I/K_{II}$ \\ \hline
			\textrm{SPK}& $1.01$&$4.79$&	$16.74$ & $4.74$	\\
			\textrm{PFD}& $1.44$ & $ 3.99$&$	14.30$ & $2.78$
		\end{tabular} 
		\vspace{0.5cm}
		\caption{Experimental results for $K_{II}$.}
		\label{tab:Gcresults}
	}
	
\end{table}

The mode II fracture toughness is defined as the critical mode II stress intensity factor at failure. Following Bahrami \cite{bahrami2020}, the mode II fracture toughness can be computed based on the failure load $F$, the geometrical dimensions of the specimen and the mode II stress intensity factor $K_{II}^{\ast}$ following
\begin{equation}
\label{eq:KIIcalc}
K_{II}= K_{II}^{\ast} \frac{F\,\sqrt{\pi\,a}}{\pi\,R\,t}\,,
\end{equation}
where $t$ is the thickness of the disk, $a$ is the notch length, and $R$ is the radius of the disk. Bahrami~\cite{bahrami2020} performed finite element analyses to determine the mode I and II stress intensity factors for different DNBD geometries. The relevant values for the selected geometry are summarized in Table \ref{tab:GcEvaluation}. Following Equation \ref{eq:KIIcalc}, the mode II fracture toughness is calculated and the average of all experiments is taken. The results are summarized in Table \ref{tab:Gcresults}. A mode II fracture toughness of 4.79 is calculated for SPK, while a lower $K_{II}$ of 3.99 is obtained for PFD. The mode I fracture toughness of SPK is taken from \cite{scholz2019}. For PFD it is calculated based on an empirical relationship stated in \cite{whittaker1992rock} and the material parameters determined in \cite{stockinger2021}. As stated in Table \ref{tab:Gcresults}, this yields a $K_{I}/K_{II}$ ratio of 2.78 for PFD, and a ratio of 4.74 for SPK. 

\subsubsection{Crack pattern and DIC analysis}\label{sec:exppatterns}
\begin{figure}[b!]
	\begin{minipage}[b]{0.75\linewidth}
		\includegraphics[width=0.850\textwidth]{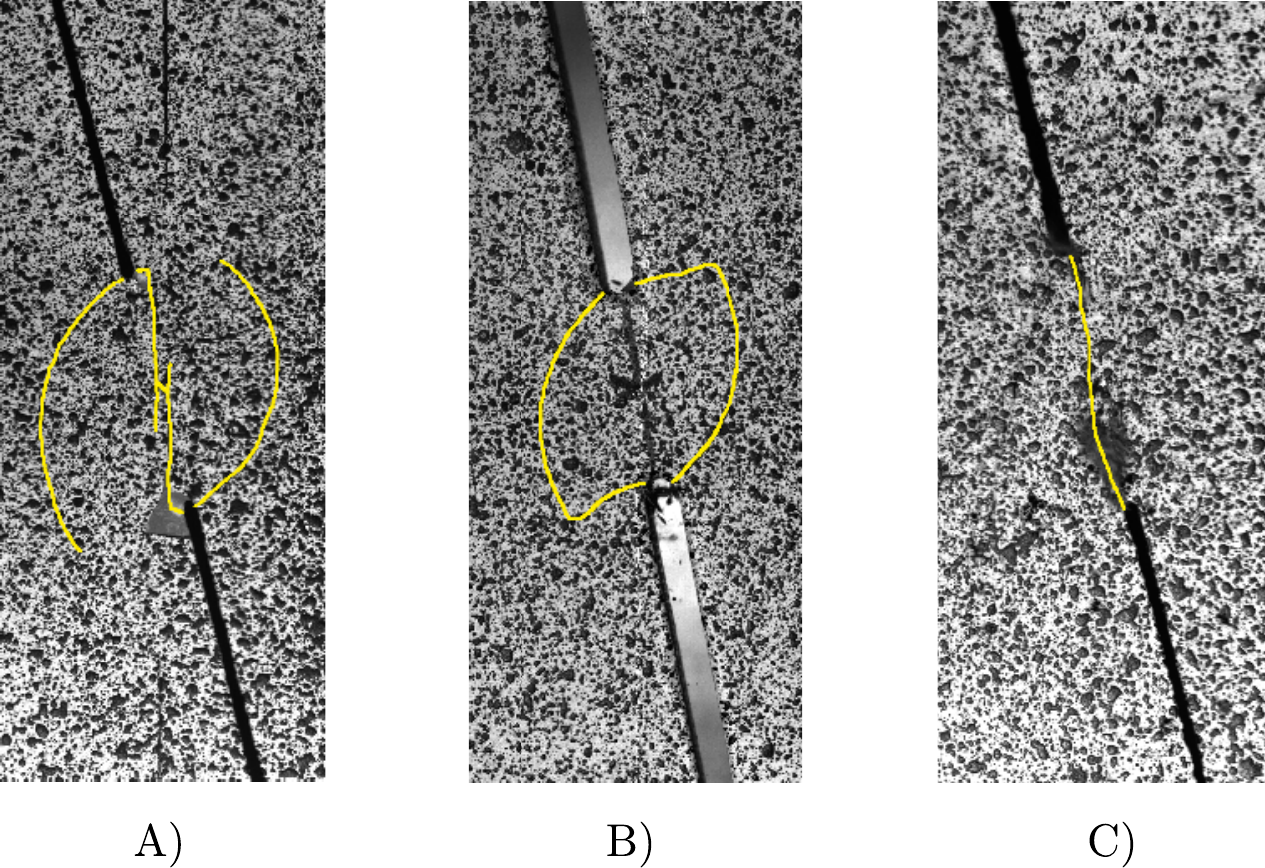}
	\end{minipage}%
	\begin{minipage}[b]{0.20\linewidth}
		\hspace{-1cm}
		\centering
		\setlength{\tabcolsep}{5pt}
		\renewcommand*{\arraystretch}{1.4}
		\begin{tabular}{l|l|l}
			Type &   SPK   & PFD \\ \hline
			A & $9$ & $7$ \\
			B & $1$ & $1$ \\ 
			C & $1$ & $3$\\  
		\end{tabular}
		\vspace{5.4cm}
	\end{minipage}
	\caption{Different crack patterns observed in the experiments: mixed tensile-shear crack (A), circular mixed tensile-shear crack (B) and pure shear crack (C) (left), and frequency of occurrence for both types of rock (right).}
	\label{fig:crackpatterns}
\end{figure}
In addition to the simple shear crack between the two notches as observed in \cite{bahrami2020}, two additional crack patterns are obtained during the DNBD tests on SPK and PFD. As shown in Figure~\ref{fig:crackpatterns}, two mixed-mode patterns are observed. Type A features two tensile wing cracks which initiate at the top and bottom notch, and emerge symmetrically. Further displacement contributes to the formation of a shear band connecting the notches, which leads to abrupt rupture of the specimen. Type B features similar tensile cracks. However, small deviations in the geometry lead to shear cracks emerging from the tips of the tensile cracks connecting to the notches. Type C is the pure shear crack observed by Bahrami \cite{bahrami2020}. In Figure~\ref{fig:crackpatterns}, right, the number of crack types observed for SPK and PFD are listed. While patterns A and B occur for SPK as well as PFD, pattern C is only observed for PFD. Consequently, both types of rock facilitate the formation of tensile wing cracks preceding the shear failure. For SPK, in all experiments two tensile wing cracks preceding the shear crack could be observed. In the case of PFD, in two out of ten experiments no tensile cracks initiated, and in four experiments only one tensile wing crack emerged. This observation fits very well to the $K_I/K_{II}$ ratios determined. SPK has a higher $K_I/K_{II}$ ratio than PFD, which explains the higher number of tensile cracks observed experimentally.

\section{Numerical results} \label{sec:results}

The phase-field model presented in Chapter \ref{sec:phasefield-theory} is calibrated for each type of rock using the computed mode II fracture toughness. In Section \ref{sec:DNBD_results} the DNBD tests are analysed. A first study (Section \ref{sec:study}) shows that the proposed model can reproduce the three crack patterns observed experimentally. To prove the correctness of the computed material parameters, specific geometries of the DNBD tests are simulated and compared with the experimental results in Section \ref{sec:validation}. As an outlook, a uniaxial compression test on a rare drill core is presented in Section \ref{sec:coreSample} .
\begin{figure}[!t]
	\centering
	\includegraphics[width=0.88\textwidth]{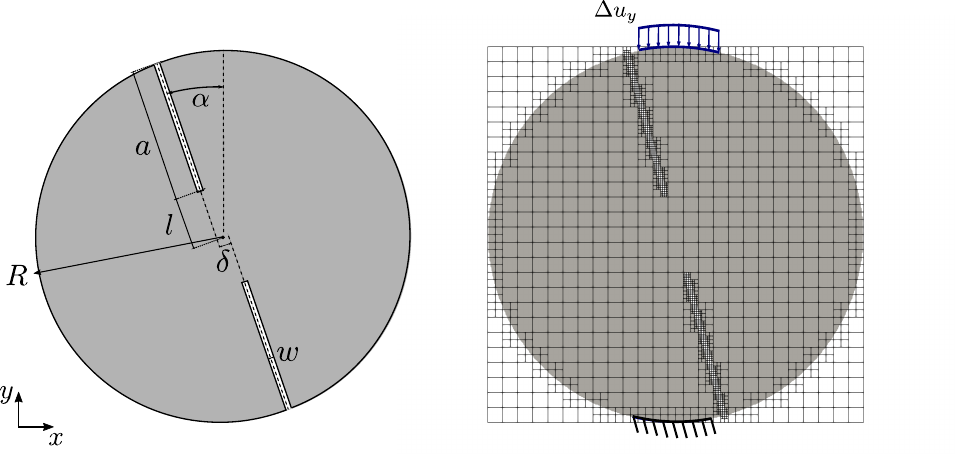}
	\caption{Setup of the computational domain (left), and mesh with initial refinement and boundary conditions (right).}%
	\label{fig:simsetupDNBD}
\end{figure}
\subsection{DNBD Experiments} \label{sec:DNBD_results}
The 2D setup for the mixed-mode simulation is shown in Figure~ \ref{fig:simsetupDNBD}, and the simulation parameters for the different crack types are listed in Table \ref{tab:params}. The computational domain is initially discretized with $25 \times 25$ elements and refined towards the specimen boundary with a refinement depth of $k=1$ and towards the notches with a refinement depth of $k=3$. An adaptive refinement strategy based on the value of the phase-field is used, which refines the mesh in each staggered step based on the criterion $s < 0.7$ up to a depth of $k=4$ for SPK, and up to a depth of $k=3$ for PFD. To resolve the geometry, a quad-tree subdivision approach with a partitioning depth of $s = 3$ is used. Dirichlet boundary conditions are applied on two circular arcs of length $10$ mm representing the contact zone between the wood and the sample on the top and, respectively, the bottom of the specimen (see Figure~ \ref{fig:simsetupDNBD}). While the displacements on the lower arc are fixed, negative displacements in the $y$-direction are applied on the top arc. An adaptive load stepping scheme is used based on the ideas presented in \cite{gupta2020}. Here, we use a step-size controller relating the step size change to the ratio of tolerance $\varepsilon$ and current error $\varepsilon_i$ which computes the load step size in iteration $i+1$ based on
\begin{flalign}
\Delta u_{i+1} = \textrm{max}\left(\Delta u_i \left( \frac{\varepsilon}{\varepsilon_i} \right)^\kappa,\, u_\textrm{min}\right)\,,&&
\end{flalign}
where $\kappa=1.1$, $u_\textrm{init}=5 \cdot 10^{-3}$  and $u_\textrm{min} = 5 \cdot 10^{-4}$. The material parameters for both types of rock are listed in Table \ref{tab:material}. While $G_{c_I}$ is obtained directly from literature in the case of Solnhofen Limestone \cite{scholz2019}, for Pfraundorfer Dolostone it is computed from $K_I$ using the relation
\begin{table}[!t]
	\centering
	\setlength{\tabcolsep}{5pt}
	\renewcommand*{\arraystretch}{1.4}
	\begin{tabular}[b]{c|c|c|c|c|c|c|c|c|c|c|c|c} 	
		$w$ [mm]& $\delta$ [mm] & $n_{x}$  &$n_{y}$ & $a_t$ [mm]   & $a_b$ [mm]  & $R$ & $\alpha $ & $p$  & $\varepsilon$ & $\beta$ & $u_\textrm{init}$  & $u_\textrm{min}$ \\ 
		\hline
		$ 1.1^{a}\:(2.7^{b})$ &$0.0^{a}\:(1.0^{b})$&$25$    & $25$   &$37.5$ & $37.5$  & $47$ & $15\degree$ & $3$ & $10^{-5}$ &$10^{6}$ & $5 \cdot 10^{-3}$ & $5 \cdot 10^{-4}$ \\
	\end{tabular}
	\caption{Choice of simulation parameters for the DNBD tests with two different geometries: $^{a}$water jet cutter and  $^{b}$hand saw.  }%
	\label{tab:params}
\end{table}
\begin{table}[b!]
	\setlength{\tabcolsep}{5pt}
	\renewcommand*{\arraystretch}{1.4}
	\begin{tabular}{lll} \hline
		
		&   SPK   & PFD \\ \hline
		$E$ [GPa] & $45.8^{\ast}$ & $52.5^{\ast}$ \\
		$\nu$ [-] & $0.31^{\ast}$ & $0.27^{\ast}$ \\ 
		$\sigma_t$ [Mpa] & $14.4^{\ast}$ & $10.9^{\ast}$ \\ 
		$G_{c_I}$ [kN/mm] & $1.97 \cdot 10^{-5\,\dagger}$ & $3.928 \cdot 10^{-5}$ \\
		$G_{c_{II}}$ [kN/mm] & $4.98 \cdot 10^{-4}$ & $3.0366 \cdot 10^{-4}$ \\
		$l_{{0,I}}$ [mm] & $0.259$ & $0.916$ \\
		$l_{{0,II}}$ [mm] & $0.682$ & $0.656$ \\
		\hline
		\vspace{0.05em}
	\end{tabular}
	\caption{Material parameters for Solnhofen Limestone and Pfraundorfer Dolostone from $^{\ast}\,$\cite{stockinger2021}, $^{\dagger}\,$\cite{scholz2019}.}%
	\label{tab:material}
\end{table}

\begin{flalign}
	G_{c_I} = \frac{K^2_I}{E}\,&&
\end{flalign}
(\cite{lawn1993fracture}). The mode II fracture energy $G_{c_{II}}$ is obtained for both rocks by inserting the experimentally determined $K_{II}$ into the analogue expression 
\begin{flalign}
G_{c_{II}} = \frac{K^2_{II}}{E}\,.&&
\end{flalign} 
Following the argumentation in \cite{tanne2018crack}, $l_0$ is treated as a material parameter. The length scale $l_{0,I}$ associated to mode I failure is determined using the tensile strength $\sigma_t$ of the material, while the length scale $l_{0,II}$ is computed based on the shear strength $\tau$, following
\begin{flalign}
	&l_{0,I} = \frac{27\,G_{c_{I}}\, E}{512\, \sigma_t^2}\,,&& \\
	&l_{0,II} = \frac{27\,G_{c_{II}}\, E}{512\, \tau^2}\,.&&
\end{flalign}
Here, we assume that the shear strength is equal to the maximum strength measured in the DNBD experiments. Consequently, we obtain different length-scale parameters for tensile and shear failure, namely $l_{{0,I}_{\,\textrm{SPK}}}=0.259$ and $l_{{0,II}_{\textrm{SPK}}}=0.682$ for SPK, and $l_{{0,I}_{\,\textrm{PFD}}}=0.916$ and $l_{{0,II}_{\textrm{PFD}}}=0.656$ for PFD.

\subsubsection{Crack patterns} \label{sec:study}
\begin{figure}[b!]
	\centering
	\includegraphics[width=0.63\textwidth]{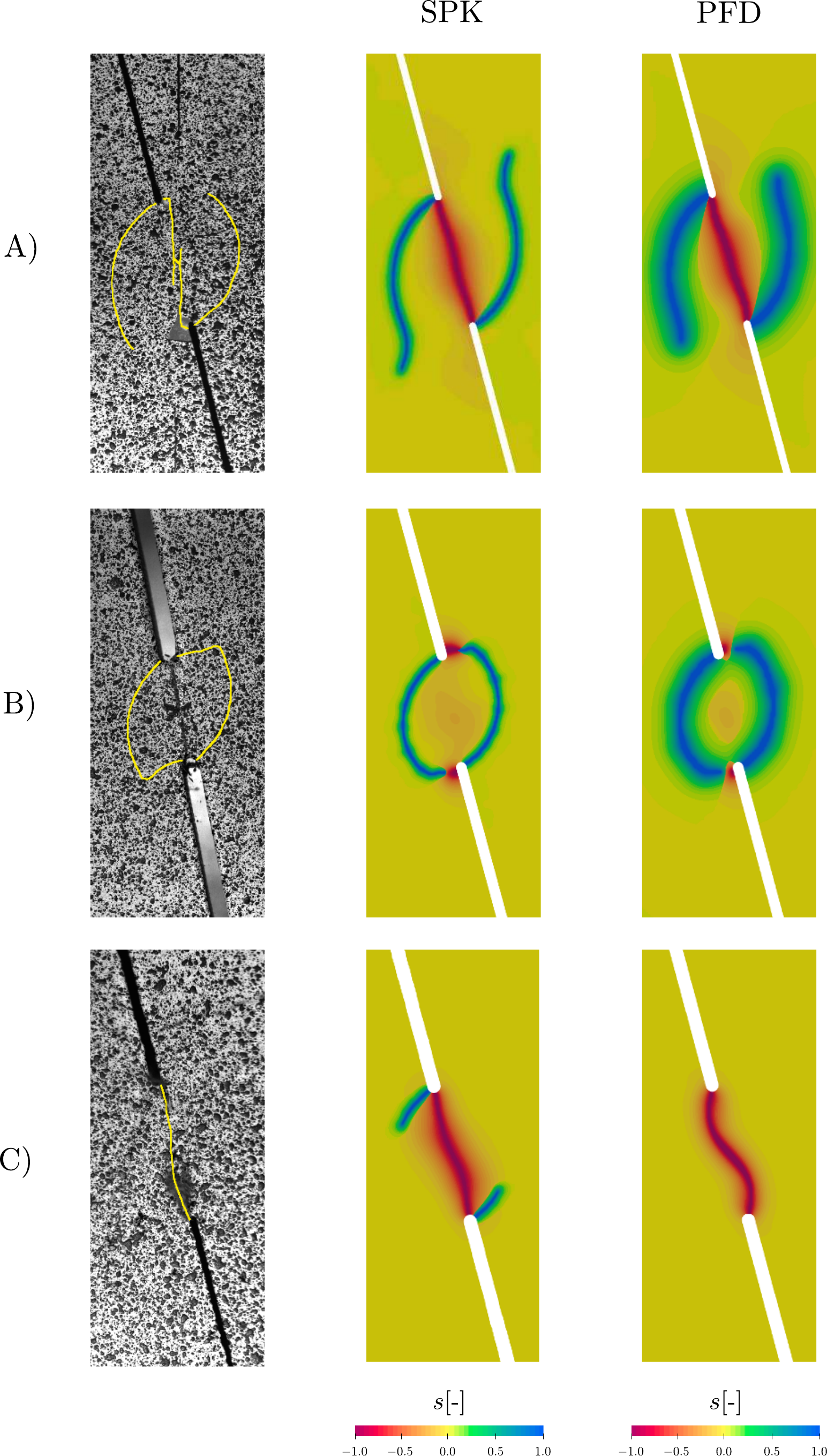}
	\caption{Comparison of experimental and simulated crack paths for the three crack types obtained for SPK and PFD: the mixed tensile-shear crack (A), the circular mixed tensile-shear crack (B) and the pure shear crack (C).}%
	\label{fig:threecracktypes}
\end{figure}
As explained in Section \ref{sec:exppatterns}, three different crack patterns could be observed in the DNBD tests: the mixed tensile-shear crack (A), the circular mixed tensile-shear crack (B) and the pure shear crack (C). To assess the possibilities of the proposed model, geometrical and material parameters are varied to see if all crack types can be reproduced. For crack types A and C, a notch width of $1.1$ mm is used, which corresponds to the geometry prepared by a water jet cutter. For Type B, a notch width of $2.0$ mm is used and the notches are shifted perpendicular to their connecting line using an offset $\delta$ of size $1.0$ mm. The latter corresponds to the geometry prepared using a hand saw. \\
The results of the phase-field simulations for both types of rock, SPK and PFD, are shown in Figure~\ref{fig:threecracktypes}. Here, for visualisation purposes a combined scale with a value of $s=1$ on a tensile crack, and a value of $s=-1$ on a shear crack is chosen. As can be seen, the proposed model is able to correctly capture all three crack types.
The mixed tensile-shear crack (A) features two tensile wing cracks emerging from the tips of the notches followed by a shear crack connecting the notch tips. This crack pattern can be reproduced for both SPK and PFD using the geometric parameters of the specimens prepared by a water jet cutter. Due to the higher $K_I/K_{II}$ ratio of SPK, the tensile wing cracks propagate further before shear failure occurs compared to PFD.
The circular mixed tensile-shear crack (B) exhibits similar tensile wing cracks, however, shear cracks develop between the tips of the tensile cracks and the respective, closest notch tips. Similar to Type A, it can be obtained using both SPK and PFD material parameters. Interestingly, the numerical results confirm that Type B occurs due to the imperfectly captured geometry when cutting the notches by hand saw. Due to the misaligned notches, the shear cracks start to initiate from the tips of the tensile wing cracks and evolve towards the notch tips, instead of the shear band forming in the center of the specimen as in the case of crack Type A. Here, the proposed model gives valuable insights concerning the formation and type of cracks, while experimental determination would require advanced experimental techniques and expensive equipment.
The pure shear crack (C) shows a shear band connecting the two notches. In contrast to Type A, no or only minor tensile wing cracks have formed before shear failure occurs. Whereas crack types A and B are triggered by geometric differences in the specimen, the pure shear crack is obtained numerically by decreasing the $K_I/K_{II}$ ratio and occurs for both geometries. For SPK, a $K_I/K_{II}$ ratio of $2.23$ is chosen, while for PFD a ratio of $2.12$ yields a pure shear crack. Whereas the numerical result for PFD represents a pure shear crack, the result obtained for SPK shows small tensile cracks which initiate before shear failure occurs. \\
In summary, the proposed model can reproduce all three experimentally observed crack patterns. A detailed study on the influence of both material and geometrical parameters on the resulting crack pattern, including stochastic analysis, might generate new valuable insights and is part of future research.
\begin{figure}[t!]
	\vspace{-0.5cm}
	\centering
	\includegraphics[height=0.9\textheight]{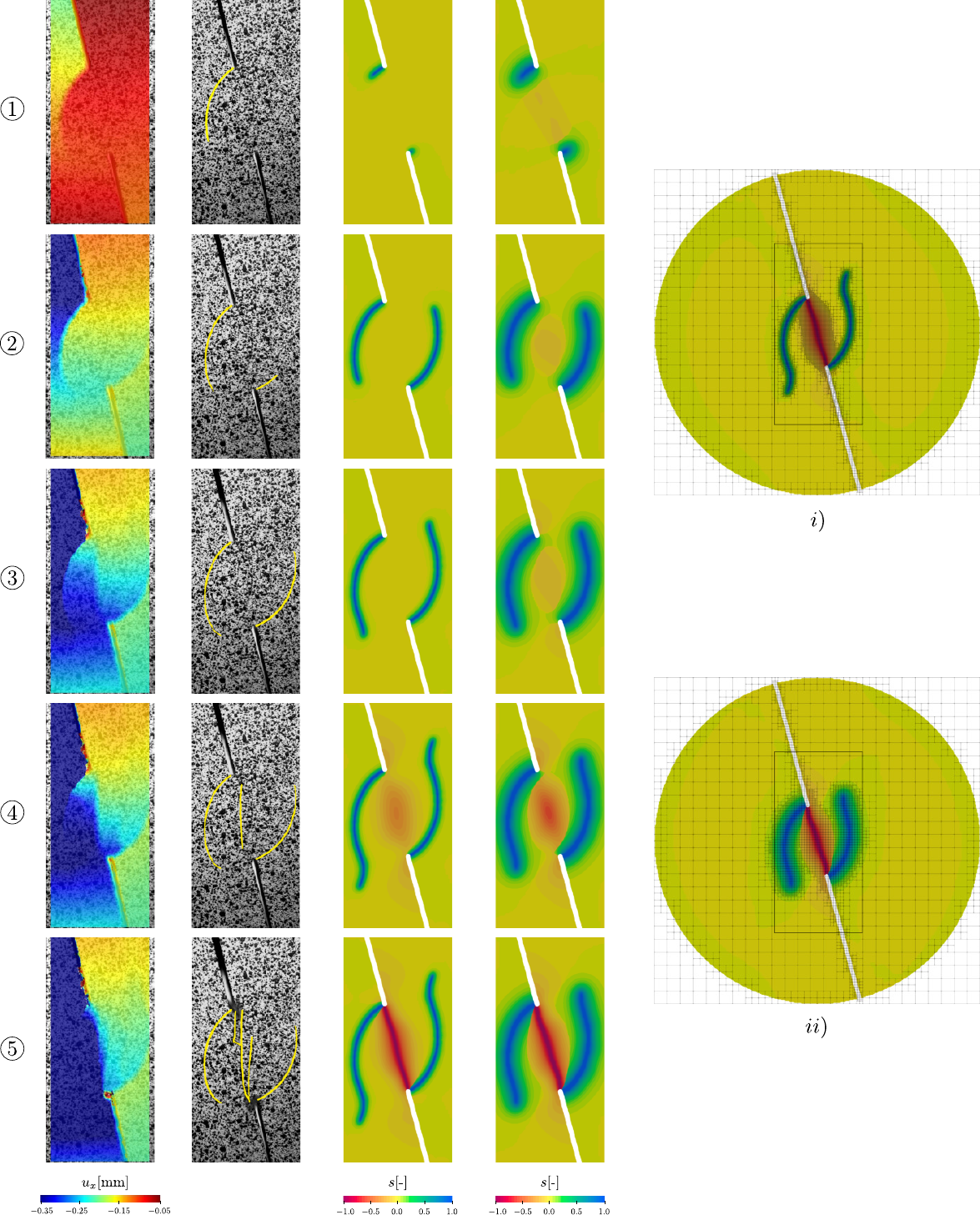}
	\caption{Comparison of experimental and numerical results for crack type A. From left to right: displacement in $x$-direction obtained by DIC, experimental crack pattern, computed phase-field for SPK and computed phase-field for PFD (left). Computation domain with mesh refinement for the fully cracked specimen for SPK (i) and PFD (ii) (right). }
	\label{fig:GcIITest_Visu}
\end{figure}
\begin{figure}[h!]
	\centering
	\includegraphics[width=0.66\textwidth]{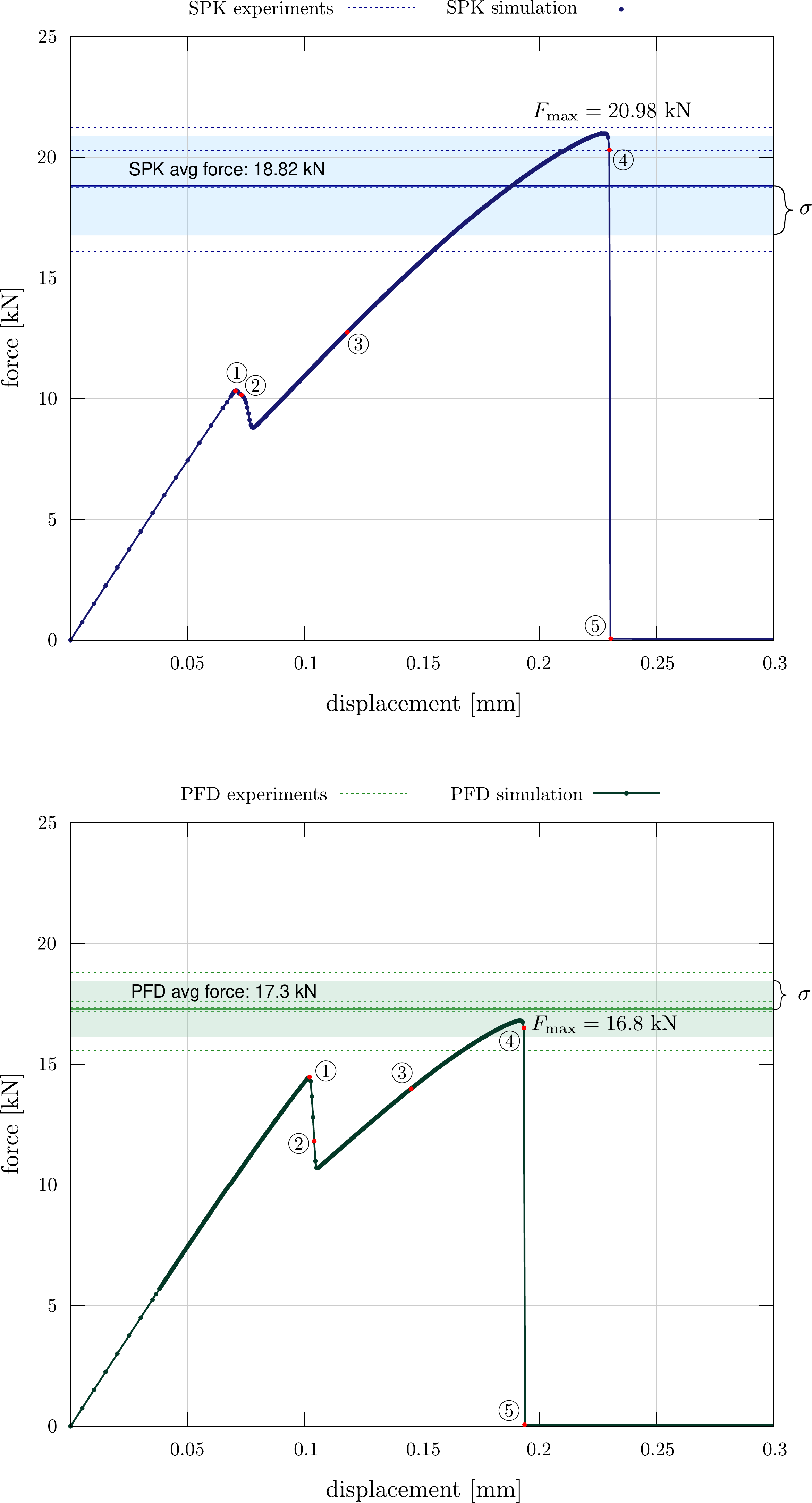}
	\caption{Computed load displacement curves for the SPK specimen (top) and PFD specimen (bottom). The points \raisebox{-0.05cm}{\includegraphics[height=0.9em]{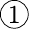}} - \raisebox{-0.05cm}{\includegraphics[height=0.9em]{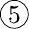}} correspond to the patters shown in Figure~\ref{fig:GcIITest_Visu}. The experimental failure loads with standard deviation and the average failure loads are shown for both types of rock. The computed failure loads yield a deviation of $11.48\%$ for SPK and a deviation of $2.98\%$ for PFD from the respective averaged experimental failure loads.}
	\label{fig:GcII_loadcurves}
\end{figure}

\subsubsection{Validation} \label{sec:validation}

In this section, a detailed analysis of the mixed tensile-shear crack (Type A) is presented. To this end, only experiments yielding this specific crack pattern are considered. This corresponds to $5$ SPK and $5$ PFD experiments of the geometry prepared using a water jet cutter, which yielded mostly Type A crack patterns.
In Figure~\ref{fig:GcIITest_Visu}, left, experimental and numerical crack patterns are compared for both types of rock. On the left, five different phases of crack propagation are evaluated for the experiments based on the high-speed camera recordings and compared with the simulation results. Here, the first column shows the displacement in the $x$-direction computed with DIC, while the second column shows the corresponding crack path. The computed phase-field crack paths are depicted in the third column (SPK) and the fourth column (PFD). 
In phase \raisebox{-0.05cm}{\includegraphics[height=0.9em]{graphics/symbol1.pdf}}, a wing crack starts to initiate at the notch where the displacement is applied. This behavior can be observed both in the experiments and in the numerical simulation. However, in the phase-field analysis, the second tensile crack initiates much earlier and not only after the first tensile crack has almost fully developed. This difference in behavior likely stems from the different boundary conditions applied in experiments and simulations. In the experiment, the disk can compress and sink into the soft wood, resulting in a change of the contact area over time. This behavior is not captured by the boundary conditions set for the numerical simulation. In phase \raisebox{-0.05cm}{\includegraphics[height=0.9em]{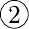}}, the propagation of the tensile cracks continues with increasing displacement. The growth of the tensile wing cracks steadily decelerates as soon as the wing cracks propagate up to the height of the opposite notch tip (phase \raisebox{-0.05cm}{\includegraphics[height=0.9em]{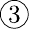}}). Next, a shear band starts to develop in the center of the disk. In contrast to the tensile cracks, which initiate locally and grow from the crack tip with increasing load, the shear failure follows a different pattern. As can be seen, the shear crack initiates at the center of the disk and the associated damage covers a wider area. The damage increases gradually along the connection line between the notches (phase \raisebox{-0.05cm}{\includegraphics[height=0.9em]{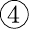}}) until failure occurs abruptly in phase \raisebox{-0.05cm}{\includegraphics[height=0.9em]{graphics/symbol5.pdf}}. \\
In summary, the numerical behavior, including the localisation of the shear band in the center of the specimen, agrees very well with the experimental observations. The fully developed crack patterns for SPK and PFD are shown in Figure~\ref{fig:GcIITest_Visu}, right. A direct comparison of the crack patterns of the two types of rock shows that the tensile wing cracks propagate further in the case of SPK. Due to the higher $K_I/K_{II}$ ratio, the initiation of shear cracks occurs later, which enables the wing cracks to propagate beyond the notch tip of the opposite notch. This difference is also visible in the experimental crack paths.
The computed load-displacement curves are shown in Figure~\ref{fig:GcII_loadcurves} for SPK, top, and PFD, bottom). Due to the different boundary conditions in the experiment no direct comparison of experimental and simulated load-displacement curves is presented here. As the plastic deformations of the wooden plates which are used to transfer the load to the specimen can not be captured by the numerical model, in the following, the computed failure loads are compared against the averaged experimental failure load. For both types of rock, the crack phases \raisebox{-0.05cm}{\includegraphics[height=0.9em]{graphics/symbol1.pdf}} - \raisebox{-0.05cm}{\includegraphics[height=0.9em]{graphics/symbol5.pdf}} are marked. First, the force increases linearly until the tensile wing cracks start to initiate (\raisebox{-0.05cm}{\includegraphics[height=0.9em]{graphics/symbol1.pdf}}). This results in a sudden drop in force which occurs at $10.18$ kN in the case of SPK and at $14.37$ kN in the case of PFD. Due to the different length scales for tensile cracks ($l_{{0,I}_{\,\textrm{SPK}}}=0.259$ and $l_{{0,I}_{\,\textrm{PFD}}}=0.916$), the loss in force is higher for PFD. The tensile wing cracks are not yet fully developed, when the force starts increasing again. The slope is almost linear until \raisebox{-0.05cm}{\includegraphics[height=0.9em]{graphics/symbol3.pdf}}, when the shear band starts to develop. Failure occurs at the maximum bearable force of $F_\textrm{max, SPK} = 20.98$ kN and $F_\textrm{max, PFD} = 16.8$ kN, respectively. At this point (\raisebox{-0.05cm}{\includegraphics[height=0.9em]{graphics/symbol4.pdf}}), the shear damage is already clearly visible. Once the shear band has fully developed (\raisebox{-0.05cm}{\includegraphics[height=0.9em]{graphics/symbol5.pdf}}), the force drops to zero. As can be seen, the computed failure loads are in very good agreement with the experimental values. For both types of rock, the computed values lie within the range of values observed experimentally. The numerical failure loads show a relative deviation of $11.48\%$ for SPK and of $2.98\%$ for PFD from the respective averaged experimental failure loads.

\subsection{Rare drill core}\label{sec:coreSample}
\begin{figure}[t!]
	\centering
	\includegraphics[width=0.95\textwidth]{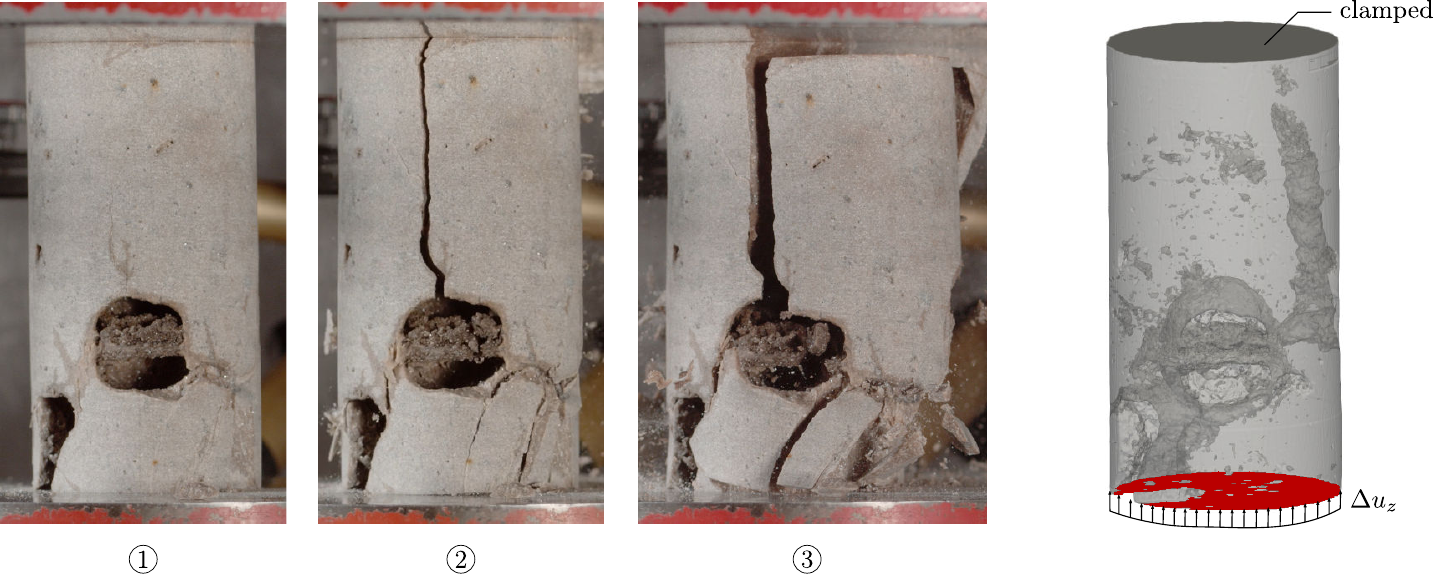}
	\caption{Experimental setup and recorded crack pattern (left), and simulation setup (right) based on the CT-scanned drill core.}
	\label{fig:setup_coreSample}
\end{figure}

\begin{table}[!b]
	\centering
	\setlength{\tabcolsep}{5pt}
	\renewcommand*{\arraystretch}{1.4}
	\begin{tabular}[b]{c|c|c|c|c|c|c|c|c|c|c|c} 	
		$n_{x}$  &$n_{y}$ & $n_{z}$   & $p$  & $\varepsilon$ & $\beta$ & $u_\textrm{large}$  & $u_\textrm{med}$ & $u_\textrm{small}$ \\ 
		\hline
		$44$    & $92$   &$44$ & $3$ & $1 \cdot 10^{-5}$ &$10^{6}$ & $2 \cdot 10^{-2}$ & $4 \cdot 10^{-3}$ & $3 \cdot 10^{-3}$\\
	\end{tabular}
	\caption{Choice of simulation parameters for the drill core. }%
	\label{tab:coreparams}
\end{table}
In this section, the proposed three-field model is applied to a complex, 3-dimensional crack scenario. Within the framework of the Geothermal-Alliance Bavaria, various laboratory experiments could be carried out on rare drill cores of the exploration well Moosburg SC 4 (MSC-4) (\cite{bohnsack2020porosity, bohnsack2021stress, potten2020geomechanical}). Drilled in $1990$ to a total vertical depth of $1585$ m, the MSC-4 well is unique for being fully cored over the entire reservoir section of the Upper Jurassic carbonates (Malm aquifer) with a thickness of $453$ m (\cite{bohm2011tafelbankiger, bohm2012lithofazies, meyer1994moosburg}). In the following, a uniaxial compression test on a rare drill core from a dolomitic part of the reservoir is presented. The low-porosity dolostone sample shows crystalline sizes of $0.25$ mm to $1.0$ mm as well as small and large vugs. To obtain the drill core's exact geometry, the cylindrical sample with a height of $98.3$ mm and a diameter of $49.7$ mm was CT-scanned with a resolution of $0.11\: \textrm{mm} \times 0.11\:  \textrm{mm} \times 0.11\:  \textrm{mm}$. A uniaxial compression test was performed using a displacement-controlled test speed using a displacement rate of $0.06$ mm/min. To be able to analyze the experimental crack pattern, the compression test was recorded using a Phantom Flex4K high-speed camera with 2000 fps in full HD resolution. \\
The experimental setup and observed crack pattern is shown in Figure~\ref{fig:setup_coreSample}, left. Upon loading, a vertical crack initiates on the upper side of the large, central pore on the front side of the specimen (Figure~\ref{fig:setup_coreSample}, \raisebox{-0.05cm}{\includegraphics[height=0.9em]{graphics/symbol1.pdf}}). In addition, cracks emerge which connect to the smaller pore on the right side of the specimen as well as the lower side of the specimen (Figure~\ref{fig:setup_coreSample}, \raisebox{-0.05cm}{\includegraphics[height=0.9em]{graphics/symbol2.pdf}}). The vertical crack continues to grow upwards until it reaches the top side of the drill core. The premature damage on the left side of the large pore leads to the development of smaller cracks which will connect to a pore on the back of the core sample, as seen in Figure~\ref{fig:setup_coreSample}, \raisebox{-0.05cm}{\includegraphics[height=0.9em]{graphics/symbol3.pdf}}. Here, failure occurs and the right half of the specimen is blasted off. The failure pattern is dominated by the pores inside the rock, which trigger the initiation of the cracks. 

\begin{figure}[b!]
	\centering
	\includegraphics[width=0.92\textwidth]{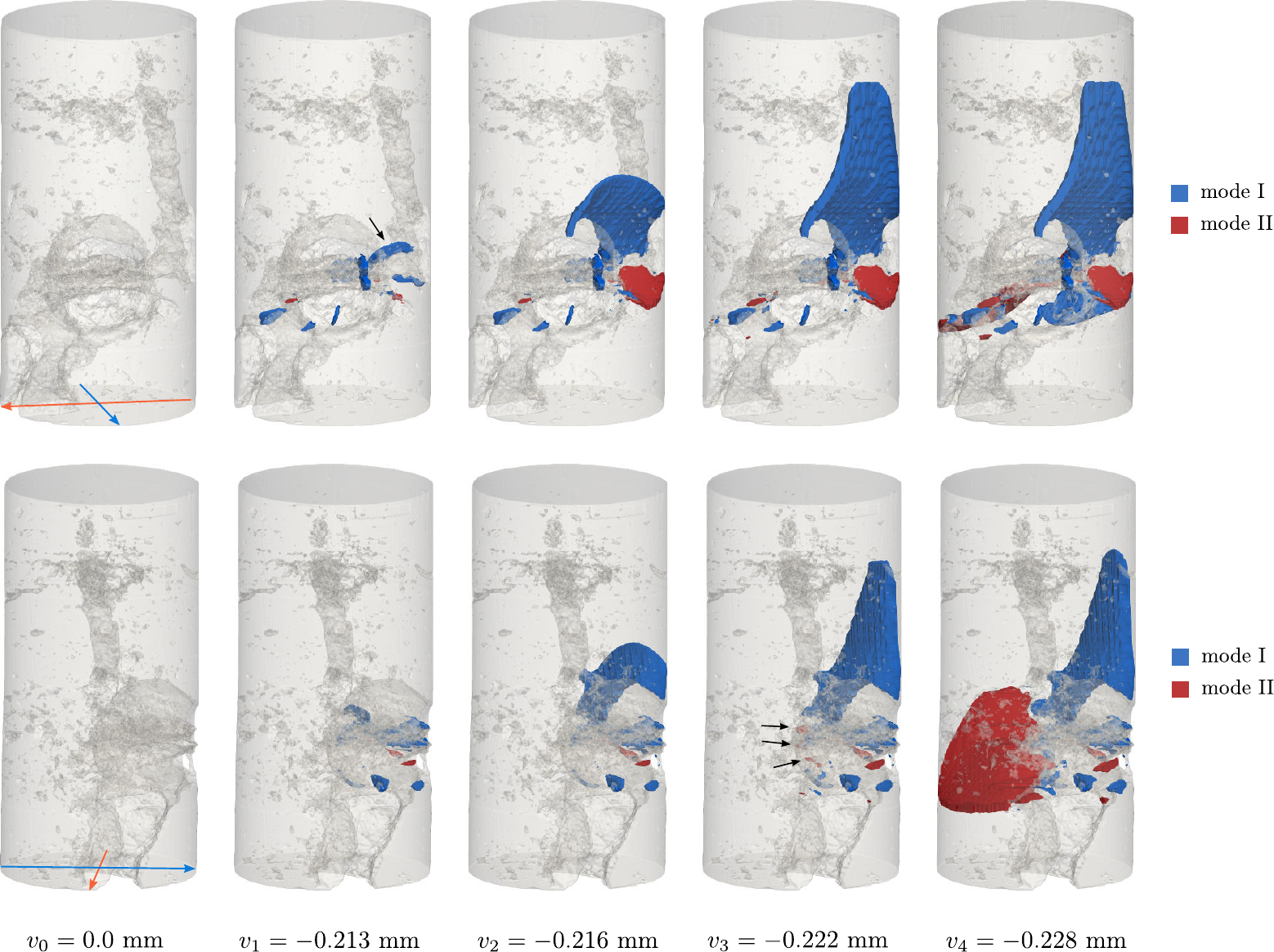}
	\caption{Simulated crack pattern for different displacement steps. For visualisation purposes the cracks are shown as iso-volumes of the phase-field, i.e. $s_I \leq 0.03$ for the tensile cracks depicted in blue, and $s_{\textrm{II}} \leq 0.03$ for the shear cracks depicted in red. Two different views are shown: a front view corresponding to the viewing angle of the camera in the experiments (top row), and a side view (bottom row). The red and orange line in the first column illustrate the orientation of the different views. }
	\label{fig:crack_coreSample}
\end{figure}
\subsubsection{Simulation Setup}
The geometry and boundary conditions for the numerical simulation are depicted in Figure~\ref{fig:setup_coreSample}, right, and the simulation parameters are listed in Table \ref{tab:coreparams}. The boundary conditions are set as follows: $y$-displacements are fixed on the top surface, while a positive displacement is applied in $y$-direction. Steps of size $0.02$ mm are applied until a total displacement of $0.2$ mm, at which the step size is decreased to $0.005$ mm. The geometry is represented using FCM. To this end, the core sample is embedded into a Cartesian mesh with $44 \times 92 \times 44$ elements and integration is performed using an octree sub-division approach with partitioning depth $s = 3$. This results in a total number of $2\,168\,105$ DOFs for the three-field system. Due to the limited access to the material of the rare drill core, the parameters for this dolostone could not be determined experimentally. Consequently, it is assumed that the MSC-4 dolostone behaves similar to the PFD and material parameters are taken from Table~\ref{tab:material}. However, the mode II fracture toughness needs to be chosen differently, as the value determined for PFD results in premature cracking and a crack pattern which does not relate to the one observed experimentally. Based on a parameter study we choose $G_{c_\textrm{II}} = 5 \cdot 10^{-3}$ kN/mm, for which we obtain good agreement in both the failure load as well as the observed crack pattern. The higher value of $G_{c_\textrm{II}}$ can be attributed to the three dimensional setting, in which not only mode I and mode II, but also mode III cracks occur. However, the extension of the proposed two-field problem to account for mode III cracks is the subject of future research and is beyond the scope of this work. The simulation is performed on the \mbox{SuperMUC-NG} cluster at the Leibniz Supercomputing Centre using 24 nodes with 48 cores and 96 GB memory per node. The hybrid MPI and OpenMP parallelisation of the three-field phase-field problem is based on the framework presented in \cite{jomo2017parallelization, jomo2019robust}. 
\begin{figure}[t!]
	\centering
	\includegraphics[width=0.65\textwidth]{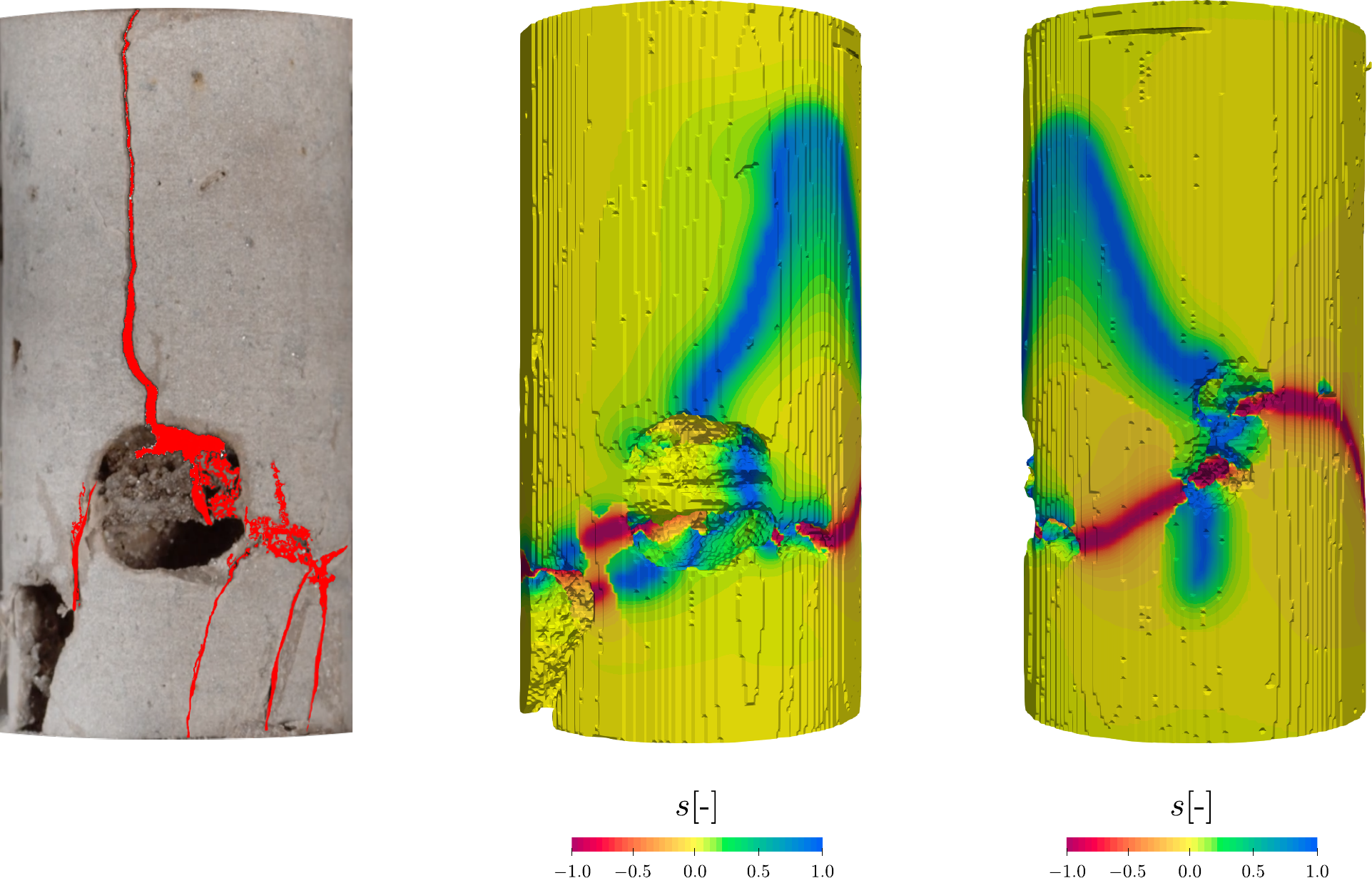}
	\caption{Comparison of the experimental crack pattern with manually highlighted cracks (left), and computed tensile and shear cracks obtained with the three-field model (right). }
	\label{fig:comp_coreSample}
\end{figure}
\subsubsection{Results}
\begin{figure}[h!]
	\centering
	\vspace{1em}
	\includegraphics[width=0.75\textwidth]{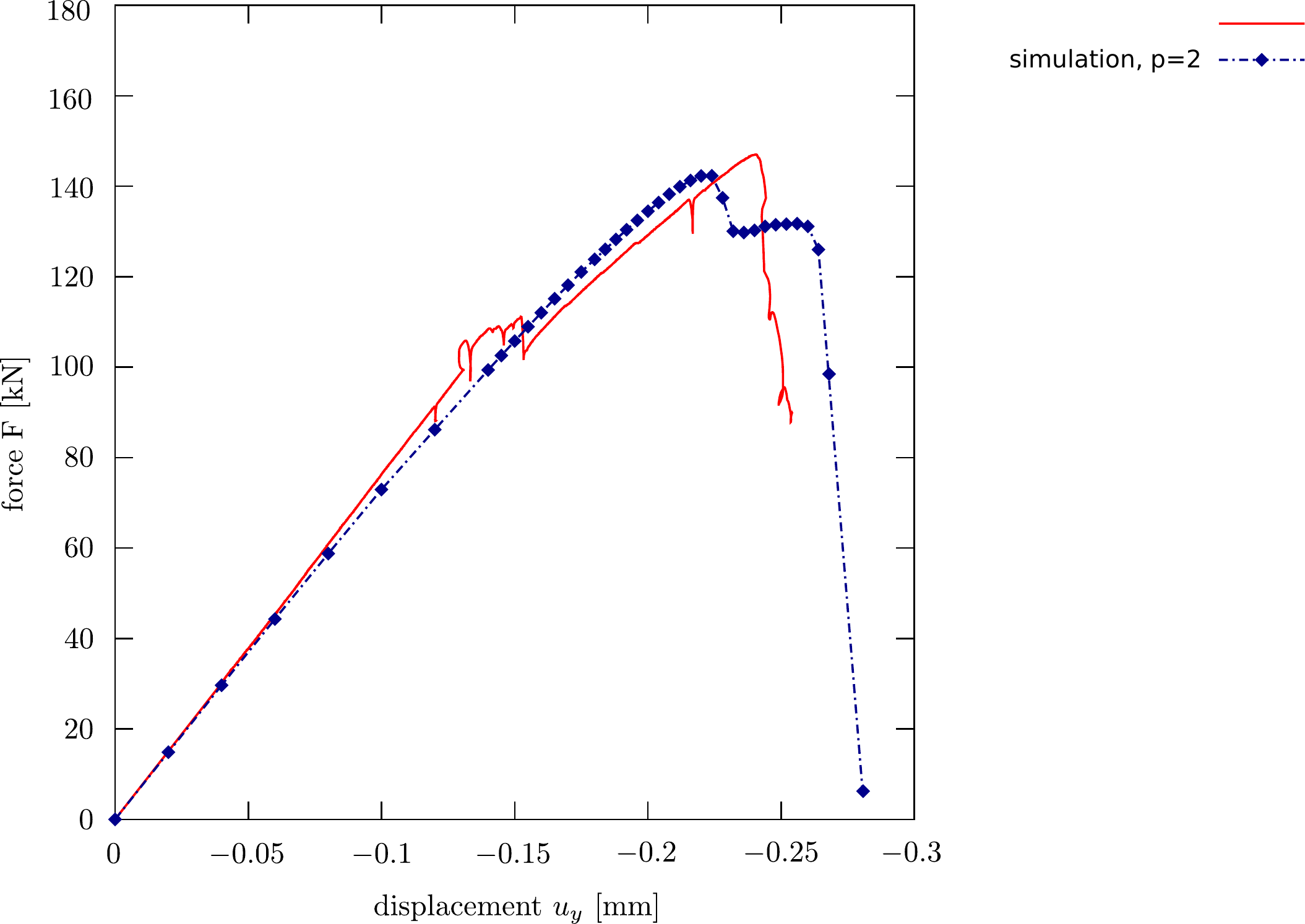}
	\caption{Comparison of the experimental and computed load displacement curves. While the experimentally measured fracture force is $147.0$ kN, the numerical simulation predicts failure at  \mbox{$142.31$ kN} with corresponds to a deviation of $3.2\: \%$.  }
	\label{fig:loaddisp_coreSample}
\end{figure}

In Figure~\ref{fig:crack_coreSample}, the computed fracture pattern is shown for different displacement steps. The tensile cracks and shear cracks are visualised as iso-volumes of their respective phase-field based on a choice of $s_I \leq 0.03$ for the tensile cracks shown in blue, and $s_{\textrm{II}} \leq 0.03$ for the shear cracks shown in red. At a displacement of $v_1=-0.22$ mm, cracks initiate around the large central pore. Due to the lower $G_c$ value for mode I fracture, most cracks are of tensile nature. Only a few shear cracks emerge from the left and right sides of the central pore directed towards the smaller pores. A tensile crack starts to initiate on an internal pore, which is highlighted with a black arrow. With further displacement, the crack propagates upwards, connecting the large, central pore with the longitudinal pore on the back of the drill core (step $v_2$). Additionally, a shear crack occurs, which initiates on the right side of the central pore and propagates towards the specimen's back, where it connects to the bottom end of the longitudinal pore. At displacement step $v_3$, the vertical tensile crack has propagated further towards the top plate of the drill core, while shear damage accumulates at the back of the central pore \mbox{(step $v_3$)}. At a displacement of $v_4=-0.268$ mm, a large shear crack has emerged from the back of the central pore leading to complete failure of the sample.\\
For a detailed comparison of the final crack pattern, the experimental cracks are highlighted in Figure~\ref{fig:comp_coreSample} and contrasted with the numerical result. A common feature is the vertical tensile crack which initiates on top of the internal pore and propagates upwards. In contrast to the experiment, where a nearly straight vertical crack is observed, the computed crack tends to lean towards the outer surface of the specimen. This behavior is related to the boundary conditions. Firstly, as can be seen in the experimental setup (cf. Figure~\ref{fig:setup_coreSample}), the top and bottom surface of the drill core are not completely parallel. This results in a real displacement which differs from the pure in-axis displacement applied in the simulation. Moreover, as a consequence of the phase-field formulation, the crack is not able to penetrate the Dirichlet boundary. Instead, it isrepelled from the top and bottom surface of the core sample. Therefore, the part on the right side of the core sample that falls off in Figure~\ref{fig:setup_coreSample}, \raisebox{-0.05cm}{\includegraphics[height=0.9em]{graphics/symbol3.pdf}}, is smaller in the simulation than in the experiment and shaped differently. The vertical cracks connecting the larger pores to the bottom side of the specimen can not be reproduced in the simulation. Similar to the experiment, shear and combined tensile-shear cracks connecting the different pores are visible outside the specimen. Since the fracture pattern could only be recorded from one side, it is difficult to judge if the final failure occurred due to a shear crack. However, the visible experimental crack pattern is captured remarkably well, and the numerical simulation generates interesting insights, including the fact that the vertical tensile crack initiates on top of the internal pore, as highlighted in Figure~\ref{fig:setup_coreSample}, step $v_1$.In Figure~\ref{fig:loaddisp_coreSample}, the experimental load-displacement curve is shown and compared with the computed result. The experimental curve clearly shows a brittle fracture behavior predicting failure of the sample at a force of $147.0$ kN and a displacement of $0.24$ mm. The computed curve closely follows the slope of the experimental curve until a displacement of $0.224$ mm when the maximum load of $142.31$ kN is reached. At this point, the tensile crack starts to develop which results in a drop in force. Once the tensile crack has stabilised, the curve slowly starts to ascend again. Shear damage accumulates which results in complete failure at a displacement of $0.256$ mm. The deviation of the computed failure load corresponds to $3.2\: \%$ of the force measured experimentally.
 
\section{Conclusion} \label{sec:conclusions}
In this contribution, a three-field phase-field model for the simulation of mixed-mode fracture in rock is presented. Separate scalar phase-field variables associated to mode I and mode II failure are introduced, and the two phase-field equations are implicitly coupled through the degradation of the material in the
elastic equation. By introducing separate length scales for the mode I and the mode II problems similar to \cite{fei2021double}, the major strength of the model lies in its ability to account for different tensile and shear strengths of the material. The framework is easy to implement and flexible, as it allows the choice of different splits and degradation functions. By clearly distinguishing between tensile and shear cracks it facilitates the analysis of complex fracture patterns.\\
To validate the three-field approach, the model was calibrated for two types of rock, Solnhofen Limestone and Pfraundorfer Dolostone. The mode II fracture toughness for each type of rock was determined experimentally using double-edge notched Brazilian disk tests. The simulations of the DNBD tests demonstrate that the proposed model can reproduce the three crack patterns observed experimentally: a mixed tensile-shear crack, a circular mixed tensile-shear crack and a pure shear crack. Moreover, the computed failure loads agree very well with the averaged experimental results with a deviation of $11.48\,\%$ for SPK and a deviation of $2.98\, \%$ for PFD. To test the applicability of the model for realistic 3D fractures of complex shaped specimen the three-field model was applied to a uniaxial compression test on a rare drill core. For a detailed analysis of the crack patterns, the experiment was recorded using a high-speed camera. The exact geometry of the dolostone sample was extracted from a CT-scan. The computed crack pattern captures the most characteristic fractures observed experimentally. The recorded load-displacement curve can be reproduced with good agreement showing a deviation in the failure load of $3.2\, \%$. The deviations can be explained by uncertainties in the boundary conditions as well as the diverging material parameters of the rare drill core. The example demonstrates the ability of the model to reproduce complex, 3-dimensional crack patterns in rock and its potential to generate valuable insights in the field of mixed-mode fracture.

\newpage
\section*{Acknowledgements} 
The authors gratefully acknowledge the funding through the joint research project Geothermal-Alliance Bavaria (GAB) by the Bavarian State Ministry of Science and the Arts (StMWK). Moreover, we would like to thank the Competence Network for Scientific High Performance Computing in Bavaria (KONWIHR) and the Gauss Centre for Supercomputing e.V. (www.gauss-centre.eu) for the financial support and computing time provided on the Linux Cluster CoolMUC-2 and on the GCS Super-computer SuperMUC-NG at Leibniz Supercomputing Centre (www.lrz.de). We also extend our gratitude to the Deutsche Forschungsgemeinschaft (DFG) for its financial support through the TUM International Graduate School of Science and Engineering (IGSSE), GSC 81. Additionally, we thank John Jomo for his work on the parallelization, without which these computations would not have been possible. We would also like to express our gratitude towards Prof. Laura de Lorenzis from ETH Zürich and Roland Kruse from the Technische Universität Braunschweig for performing the CT Scans of the rare core sample free of charge in the true spirit of help among colleagues. Thank you!

\section*{Conflict of interest} 
No potential conflict of interest was reported by the authors.

\newpage
\bibliographystyle{apalike}
 \bibliography{library.bib}

\begin{thebibliography}{}

\bibitem[Aldakheel et~al., 2021]{aldakheel2021global}
Aldakheel, F., Noii, N., Wick, T., and Wriggers, P. (2021).
\newblock A global--local approach for hydraulic phase-field fracture in
  poroelastic media.
\newblock {\em Computers \& Mathematics with Applications}, 91:99--121.

\bibitem[Ambati et~al., 2015a]{ambati2015phase}
Ambati, M., Gerasimov, T., and De~Lorenzis, L. (2015a).
\newblock Phase-field modeling of ductile fracture.
\newblock {\em Computational Mechanics}, 55(5):1017--1040.

\bibitem[Ambati et~al., 2015b]{ambati2015review}
Ambati, M., Gerasimov, T., and Lorenzis, L.~D. (2015b).
\newblock A review on phase-field models of brittle fracture and a new fast
  hybrid formulation.
\newblock {\em Computational Mechanics}, 55(2):383--405.

\bibitem[Amor et~al., 2009]{amor2009regularized}
Amor, H., Marigo, J.-J., and Maurini, C. (2009).
\newblock Regularized formulation of the variational brittle fracture with
  unilateral contact: Numerical experiments.
\newblock 57:1209--1229.

\bibitem[Andreev, 1991]{andreev1991review}
Andreev, G. (1991).
\newblock A review of the brazilian test for rock tensile strength
  determination. part i: calculation formula.
\newblock {\em Mining Science and Technology}, 13(3):445--456.

\bibitem[Backers et~al., 2002]{backers2002}
Backers, T., Stephansson, O., and Rybacki, E. (2002).
\newblock Rock fracture toughness testing in mode ii—punch-through shear
  test.
\newblock {\em International Journal of Rock Mechanics and Mining Sciences},
  39(6):755--769.

\bibitem[Bahrami et~al., 2020]{bahrami2020}
Bahrami, B., Nejati, M., Ayatollahi, M.~R., and Driesner, T. (2020).
\newblock Theory and experiment on true mode ii fracturing of rocks.
\newblock {\em Engineering Fracture Mechanics}, 240:107314.

\bibitem[Bieniawski and Hawkes, 1978]{bieniawski1978suggested}
Bieniawski, Z. and Hawkes, I. (1978).
\newblock Suggested methods for determining tensile strength of rock materials.
\newblock {\em International Journal of Rock Mechanics and Mining Sciences},
  15(3):99--103.

\bibitem[Blaber, 2021]{blaber2021}
Blaber, J. (2021).
\newblock Ncorr.

\bibitem[Bleyer and Alessi, 2018]{bleyer2018}
Bleyer, J. and Alessi, R. (2018).
\newblock Phase-field modeling of anisotropic brittle fracture including
  several damage mechanisms.
\newblock {\em Computer Methods in Applied Mechanics and Engineering},
  336:213--236.

\bibitem[B{\"o}hm, 2012]{bohm2012lithofazies}
B{\"o}hm, F. (2012).
\newblock {\em Die Lithofazies des Oberjura (Malm) im Gro{\ss}raum M{\"u}nchen
  und deren Einfluss auf die tiefengeothermische Nutzung}.
\newblock Dissertation.

\bibitem[B{\"o}hm et~al., 2011]{bohm2011tafelbankiger}
B{\"o}hm, F., Birner, J., Steiner, U., Koch, R., Sobott, R., Schneider, M., and
  Wang, A. (2011).
\newblock Tafelbankiger dolomit in der kernbohrung moosburg sc4: {Ein
  Schl{\"u}ssel zum Verst{\"a}ndnis der Zuflussraten in Geothermiebohrungen des
  Malmaquifers} ({{\"O}stliches Molasse-Becken, Malm S{\"u}ddeutschland)}.
\newblock {\em Z. Geol. Wissenschaft}, 39:117--157.

\bibitem[Bohnsack et~al., 2021]{bohnsack2021stress}
Bohnsack, D., Potten, M., Freitag, S., Einsiedl, F., and Zosseder, K. (2021).
\newblock Stress sensitivity of porosity and permeability under varying
  hydrostatic stress conditions for different carbonate rock types of the
  geothermal malm reservoir in southern germany.
\newblock {\em Geothermal Energy}, 9(1):1--59.

\bibitem[Bohnsack et~al., 2020]{bohnsack2020porosity}
Bohnsack, D., Potten, M., Pfrang, D., Wolpert, P., and Zosseder, K. (2020).
\newblock Porosity-permeability relationship derived from upper jurassic
  carbonate rock cores to assess the regional hydraulic matrix properties of
  the malm reservoir in the south german molasse basin.
\newblock {\em Geothermal Energy}, 8(1):1--47.

\bibitem[Bourdin et~al., 2000]{bourd2000}
Bourdin, B., Francfort, G.~A., and Marigo, J.-J. (2000).
\newblock Numerical experiments in revisited brittle fracture.
\newblock {\em Journal of the Mechanics and Physics of Solids}, 48(4):797--826.

\bibitem[Bourdin et~al., 2008]{bour2008}
Bourdin, B., Francfort, G.~A., and Marigo, J.-J. (2008).
\newblock The variational approach to fracture.
\newblock {\em Journal of elasticity}, 91(1-3):5--148.

\bibitem[Bourdin et~al., 2011]{bourdin2011time}
Bourdin, B., Larsen, C.~J., and Richardson, C.~L. (2011).
\newblock A time-discrete model for dynamic fracture based on crack
  regularization.
\newblock {\em International journal of fracture}, 168(2):133--143.

\bibitem[Bryant and Sun, 2018]{bryant2018mixed}
Bryant, E.~C. and Sun, W. (2018).
\newblock A mixed-mode phase field fracture model in anisotropic rocks with
  consistent kinematics.
\newblock {\em Computer Methods in Applied Mechanics and Engineering},
  342:561--584.

\bibitem[D{\"u}ster et~al., 2008]{duster2008finite}
D{\"u}ster, A., Parvizian, J., Yang, Z., and Rank, E. (2008).
\newblock The finite cell method for three-dimensional problems of solid
  mechanics.
\newblock {\em Computer methods in applied mechanics and engineering},
  197(45-48):3768--3782.

\bibitem[Fairhurst and Hudson, 1999]{fairhurst1999DraftIS}
Fairhurst, C. and Hudson, J.~A. (1999).
\newblock Draft isrm suggested method for the complete stress-strain curve for
  intact rock in uniaxial compression.
\newblock {\em International Journal of Rock Mechanics and Mining Sciences \&
  Geomechanics Abstracts}, 36:279--289.

\bibitem[Fan et~al., 2021]{fan2021quasi}
Fan, M., Jin, Y., and Wick, T. (2021).
\newblock A quasi-monolithic phase-field description for mixed-mode fracture
  using predictor--corrector mesh adaptivity.
\newblock {\em Engineering with Computers}, pages 1--25.

\bibitem[Fei and Choo, 2021]{fei2021double}
Fei, F. and Choo, J. (2021).
\newblock Double-phase-field formulation for mixed-mode fracture in rocks.
\newblock {\em Computer Methods in Applied Mechanics and Engineering},
  376:113655.

\bibitem[Francfort and Marigo, 1998]{fran1998revisiting}
Francfort, G.~A. and Marigo, J.-J. (1998).
\newblock Revisiting brittle fracture as an energy minimization problem.
\newblock {\em Journal of the Mechanics and Physics of Solids},
  46(8):1319--1342.

\bibitem[Freddi and Royer-Carfagni, 2010]{freddi2010regularized}
Freddi, F. and Royer-Carfagni, G. (2010).
\newblock Regularized variational theories of fracture: a unified approach.
\newblock {\em Journal of the Mechanics and Physics of Solids},
  58(8):1154--1174.

\bibitem[Gerasimov and De~Lorenzis, 2019]{gerasimov2018penalization}
Gerasimov, T. and De~Lorenzis, L. (2019).
\newblock On penalization in variational phase-field models of brittle
  fracture.
\newblock {\em Computer Methods in Applied Mechanics and Engineering},
  354:990--1026.

\bibitem[Gupta et~al., 2020]{gupta2020}
Gupta, A., Krishnan, U.~M., Chowdhury, R., and Chakrabarti, A. (2020).
\newblock An auto-adaptive sub-stepping algorithm for phase-field modeling of
  brittle fracture.
\newblock {\em Theoretical and Applied Fracture Mechanics}, 108:102622.

\bibitem[Hansen-Dörr et~al., 2019]{hansendorr2018}
Hansen-Dörr, A.~C., {de Borst}, R., Hennig, P., and Kästner, M. (2019).
\newblock Phase-field modelling of interface failure in brittle materials.
\newblock {\em Computer Methods in Applied Mechanics and Engineering},
  346:25--42.

\bibitem[Hug et~al., 2020]{hug2020}
Hug, L., Kollmannsberger, S., Yosibash, Z., and Rank, E. (2020).
\newblock A 3d benchmark problem for crack propagation in brittle fracture.
\newblock {\em Computer Methods in Applied Mechanics and Engineering},
  364:112905.

\bibitem[Jomo et~al., 2019]{jomo2019robust}
Jomo, J.~N., de~Prenter, F., Elhaddad, M., D'Angella, D., Verhoosel, C.~V.,
  Kollmannsberger, S., Kirschke, J.~S., N{\"u}bel, V., van Brummelen, E., and
  Rank, E. (2019).
\newblock Robust and parallel scalable iterative solutions for large-scale
  finite cell analyses.
\newblock {\em Finite Elements in Analysis and Design}, 163:14--30.

\bibitem[Jomo et~al., 2017]{jomo2017parallelization}
Jomo, J.~N., Zander, N., Elhaddad, M., {\"O}zcan, A., Kollmannsberger, S.,
  Mundani, R.-P., and Rank, E. (2017).
\newblock Parallelization of the multi-level hp-adaptive finite cell method.
\newblock {\em Computers \& Mathematics with Applications}, 74(1):126--142.

\bibitem[Kuruppu et~al., 2014]{kuruppu2014isrm}
Kuruppu, M.~D., Obara, Y., Ayatollahi, M.~R., Chong, K., and Funatsu, T.
  (2014).
\newblock Isrm-suggested method for determining the mode i static fracture
  toughness using semi-circular bend specimen.
\newblock {\em Rock Mechanics and Rock Engineering}, 47(1):267--274.

\bibitem[Lawn, 1993]{lawn1993fracture}
Lawn, B. (1993).
\newblock {\em Fracture of brittle solids}.
\newblock Cambridge university press.

\bibitem[Meyer, 1994]{meyer1994moosburg}
Meyer, R. (1994).
\newblock Moosburg 4, die erste kernbohrung durch den malm unter der
  bayerischen molasse.
\newblock {\em Erlanger geologische Abhandlungen}, 123:51--81.

\bibitem[Miehe et~al., 2010]{miehe2010thermodynamically}
Miehe, C., Welschinger, F., and Hofacker, M. (2010).
\newblock Thermodynamically consistent phase-field models of fracture:
  Variational principles and multi-field fe implementations.
\newblock {\em International Journal for Numerical Methods in Engineering},
  83(10):1273--1311.

\bibitem[Mo{\"e}s et~al., 1999]{moes1999finite}
Mo{\"e}s, N., Dolbow, J., and Belytschko, T. (1999).
\newblock A finite element method for crack growth without remeshing.
\newblock {\em International journal for numerical methods in engineering},
  46(1):131--150.

\bibitem[Mutschler, 2004]{mutschler2004}
Mutschler, T. (2004).
\newblock Neufassung der empfehlung nr. 1 des arbeitskreises "versuchstechnik
  fels" der deutschen gesellschaft f{\"u}r geotechnik e. v.: Einaxiale
  druckversuche an zylindrischen gesteinspr{\"u}fk{\"o}rpern.
\newblock {\em Bautechnik}, 81:825--834.

\bibitem[Nagaraja et~al., 2019]{nagaraja2019phase}
Nagaraja, S., Elhaddad, M., Ambati, M., Kollmannsberger, S., De~Lorenzis, L.,
  and Rank, E. (2019).
\newblock Phase-field modeling of brittle fracture with multi-level hp-fem and
  the finite cell method.
\newblock {\em Computational mechanics}, 63(6):1283--1300.

\bibitem[Ortiz and Pandolfi, 1999]{ortiz1999finite}
Ortiz, M. and Pandolfi, A. (1999).
\newblock Finite-deformation irreversible cohesive elements for
  three-dimensional crack-propagation analysis.
\newblock {\em International journal for numerical methods in engineering},
  44(9):1267--1282.

\bibitem[Parvizian et~al., 2007]{parvizian2007finite}
Parvizian, J., D{\"u}ster, A., and Rank, E. (2007).
\newblock Finite cell method.
\newblock {\em Computational Mechanics}, 41(1):121--133.

\bibitem[Potten, 2020]{potten2020geomechanical}
Potten, M. (2020).
\newblock {\em Geomechanical characterization of sedimentary and crystalline
  geothermal reservoirs}.
\newblock PhD thesis, Technische Universit{\"a}t M{\"u}nchen.

\bibitem[Rao et~al., 2003]{rao2003}
Rao, Q., Sun, Z., Stephansson, O., Li, C., and Stillborg, B. (2003).
\newblock Shear fracture (mode ii) of brittle rock.
\newblock {\em International Journal of Rock Mechanics and Mining Sciences},
  40(3):355--375.

\bibitem[Scholz, 2019]{scholz2019}
Scholz, C.~H. (2019).
\newblock {\em Brittle fracture of rock}, page 1–42.
\newblock Cambridge University Press, 3 edition.

\bibitem[Steinke and Kaliske, 2019]{steinke2019phase}
Steinke, C. and Kaliske, M. (2019).
\newblock A phase-field crack model based on directional stress decomposition.
\newblock {\em Computational Mechanics}, 63(5):1019--1046.

\bibitem[Stockinger, 2021]{stockinger2021}
Stockinger, G. (2021).
\newblock {\em Fracturing in Deep Boreholes}.
\newblock Springer theses, Technische Universität München, Berlin.

\bibitem[Strobl and Seelig, 2015]{strobl2015novel}
Strobl, M. and Seelig, T. (2015).
\newblock A novel treatment of crack boundary conditions in phase field models
  of fracture.
\newblock {\em Pamm}, 15(1):155--156.

\bibitem[Tan et~al., 2015]{tan2015brazilian}
Tan, X., Konietzky, H., Fr{\"u}hwirt, T., and Dan, D.~Q. (2015).
\newblock Brazilian tests on transversely isotropic rocks: laboratory testing
  and numerical simulations.
\newblock {\em Rock Mechanics and Rock Engineering}, 48(4):1341--1351.

\bibitem[Tann{\'e} et~al., 2018]{tanne2018crack}
Tann{\'e}, E., Li, T., Bourdin, B., Marigo, J.-J., and Maurini, C. (2018).
\newblock Crack nucleation in variational phase-field models of brittle
  fracture.
\newblock {\em Journal of the Mechanics and Physics of Solids}, 110:80--99.

\bibitem[Teichtmeister et~al., 2017]{teichtmeister2017phase}
Teichtmeister, S., Kienle, D., Aldakheel, F., and Keip, M.-A. (2017).
\newblock Phase field modeling of fracture in anisotropic brittle solids.
\newblock {\em International Journal of Non-Linear Mechanics}, 97:1--21.

\bibitem[Thuro et~al., 2019]{thuro2019}
Thuro, K., Zosseder, K., Bohnsack, D., Heine, F., Konrad, F., Mraz, E., and
  Stockinger, G. (2019).
\newblock Abschlussbericht zu den arbeitspaketen der technischen universität
  münchen zum verbundprojekt: Dolomitkluft - erschließung, test und analyse
  des ersten kluftdominierten dolomitaquifers im tiefen malm des molassebeckens
  zur erhöhung der erfolgsaussichten: Teilprojekt b: Geomechanische und
  hydro-geologische parametrisierung und modellierung.

\bibitem[Ulmer et~al., 2013]{ulmer2013phase}
Ulmer, H., Hofacker, M., and Miehe, C. (2013).
\newblock Phase field modeling of brittle and ductile fracture.
\newblock {\em PAMM}, 13(1):533--536.

\bibitem[Wei et~al., 2017]{wei2017experimental}
Wei, M.-D., Dai, F., Xu, N.-W., Zhao, T., and Liu, Y. (2017).
\newblock An experimental and theoretical assessment of semi-circular bend
  specimens with chevron and straight-through notches for mode i fracture
  toughness testing of rocks.
\newblock {\em International Journal of Rock Mechanics and Mining Sciences},
  99:28--38.

\bibitem[Whittaker et~al., 1992]{whittaker1992rock}
Whittaker, B.~N., Singh, R.~N., and Sun, G. (1992).
\newblock Rock fracture mechanics. principles, design and applications.

\bibitem[Zander et~al., 2015]{zander2015multi}
Zander, N., Bog, T., Kollmannsberger, S., Schillinger, D., and Rank, E. (2015).
\newblock Multi-level hp-adaptivity: high-order mesh adaptivity without the
  difficulties of constraining hanging nodes.
\newblock {\em Computational Mechanics}, 55(3):499--517.

\bibitem[Zhang et~al., 2019]{zhang2019phase}
Zhang, P., Hu, X., Bui, T.~Q., and Yao, W. (2019).
\newblock Phase field modeling of fracture in fiber reinforced composite
  laminate.
\newblock {\em International Journal of Mechanical Sciences}, 161:105008.

\bibitem[Zhang et~al., 2017]{zhang2017modification}
Zhang, X., Sloan, S.~W., Vignes, C., and Sheng, D. (2017).
\newblock A modification of the phase-field model for mixed mode crack
  propagation in rock-like materials.
\newblock {\em Computer Methods in Applied Mechanics and Engineering},
  322:123--136.

\bibitem[Zhou et~al., 2018]{zhou2018phase}
Zhou, S., Zhuang, X., and Rabczuk, T. (2018).
\newblock A phase-field modeling approach of fracture propagation in
  poroelastic media.
\newblock {\em Engineering Geology}, 240:189--203.

\end{thebibliography}

\end{document}